\documentclass[lettersize,journal]{IEEEtran}
\usepackage{amsmath,amsfonts}
\usepackage{algorithmic}
\usepackage{algorithm}
\usepackage{array}
\usepackage[caption=false,font=normalsize,labelfont=sf,textfont=sf]{subfig}
\usepackage{textcomp}
\usepackage{stfloats}
\usepackage{float}
\usepackage{url}
\usepackage{verbatim}
\usepackage{graphicx}
\usepackage{cite}
\usepackage{hyperref}
\usepackage{booktabs}
\usepackage{longtable}
\usepackage{multirow}
\usepackage{makecell}
\begin{document}

\title{ITP-STDP: A Hardware-Efficient Intrinsic-Timing Power-of-Two Synaptic Learning Engine for On-Chip SNNs}

\author{\IEEEauthorblockN{Haihang Xia, Xinyu Zhao, Xuecheng Wang, John Goodenough,~\IEEEmembership{Member,~IEEE,} Charith Abhayaratne,~\IEEEmembership{Senior Member,~IEEE,} Panagiotis A. Panagiotou~\IEEEmembership{Member,~IEEE,} Chunyi Song,~\IEEEmembership{Member,~IEEE,} and Tiantai Deng}
\thanks{Haihang Xia, Xinyu Zhao, Xuecheng Wang, John Goodenough, Charith Abhayaratne, and Panagiotis A. Panagiotou are with the School of Electrical and Electronic Engineering, The University of Sheffield, S1 3JD Sheffield, U.K.

Chunyi Song is with Donghai Laboratory, Zhoushan 316021, China, also with the Engineering Research Center of Oceanic Sensing Technologyand Equipment, Ministry of Education, Zhoushan 316021, China, and also with the State Key Laboratory of Ocean Sensing and Ocean College, Zhejiang University, Zhoushan 316021, China.

Tiantai Deng is with Donghai Laboratory, Zhoushan 316021, China.
}
}

\markboth{Journal of \LaTeX\ Class Files,~Vol.~14, No.~8, August~2021}%
{Shell \MakeLowercase{\textit{et al.}}: A Sample Article Using IEEEtran.cls for IEEE Journals}


\maketitle

\begin{abstract}
Spiking neural networks (SNNs) have the potential to emerge as the third generation of neural networks and have attracted increasing attention across a wide range of applications. However, the large number of synaptic connections in SNNs leads to intensive weight-update computation by on-chip learning algorithms during training, resulting in substantial hardware resource utilization and energy consumption. Among existing SNN learning algorithms, spike-timing-dependent plasticity (STDP) is one of the most extensively studied and widely adopted, serving as a fundamental learning rule in SNNs. 
To address the hardware and energy overheads associated with SNN training, this paper presents an intrinsic-timing power-of-two STDP (ITP-STDP) learning engine that integrates the proposed ITP-STDP learning rule with an approximate LIF neuron model.
The proposed design is evaluated through a dedicated mean-field synaptic drift model for dynamical analysis and further validated across SNN networks of different scales and datasets. It is further implemented on both ASIC and FPGA platforms and compared with state-of-the-art approaches, including the original STDP and more complex STDP variants. The results demonstrate superior energy efficiency, higher operating speed, and lower hardware resource utilization, as the proposed design eliminates most of the computational overhead of STDP through both algorithmic and hardware-level optimizations. On the FPGA platform, the proposed design improves energy efficiency by 4.17$\times$ to 87.77$\times$ over the compared designs. On the ASIC platform, 
the proposed design achieves a 5.65$\times$ to 8.70$\times$ speedup while consuming only 1.67\% to 4.52\% of the area required by prior works.
\end{abstract}


\section{Introduction}
\IEEEPARstart{N}{eural} networks have developed rapidly in recent years, while their growing energy consumption has become a key limitation \cite{joo22}. Spiking neural networks (SNNs) provide a low-power alternative to conventional neural networks due to their event-driven characteristics \cite{liu22} and are considered the third generation of neural networks \cite{maa97}. As a result, SNNs have been widely studied in applications such as communications, robotics, and aerospace \cite{che23,nic13,eqp25}.


SNNs are characterized by biologically inspired network structures, neuron models, and learning mechanisms \cite{liu24}. Their event-driven spike communication and temporal neural dynamics make them a major computational paradigm in neuromorphic computing \cite{wan22}. However, achieving both high resource efficiency and biologically plausible plasticity in neuromorphic hardware remains challenging, owing to the complexity of the neuron and synapse models required to reproduce biological learning mechanisms \cite{fre19}. At the neuron level, the leaky integrate-and-fire (LIF) model is widely adopted in neuromorphic processors \cite{kim24}, such as Loihi \cite{dav18}, FireFly \cite{li23}, and NEXUS \cite{sad25}, due to its computational simplicity and biological plausibility \cite{xia25}. 

In terms of learning mechanisms, spike-timing-dependent plasticity (STDP) is one of the most widely studied and adopted learning rules, serving as a fundamental component of SNN learning \cite{lob19}. It builds upon Hebbian learning \cite{heb05}, which describes synaptic plasticity based on the correlation between neuronal activities. 
As a spike-based and temporally asymmetric extension of Hebbian learning, STDP governs synaptic modifications according to the precise relative timing of presynaptic and postsynaptic spikes \cite{jia25}, thereby reinforcing synaptic connections associated with causal spike interactions while weakening noncausal or anti-causal ones \cite{son00}. 

This local, timing-dependent learning mechanism enables event-driven online synaptic adaptation using only locally available pre- and postsynaptic spike events, thereby exploiting temporal sparsity to reduce unnecessary computation without requiring global error propagation \cite{voh23}.
While conventional backpropagation-based learning is effective for task-oriented optimization, it relies on task-level error information as well as gradient computation and propagation across the network, making fully local and event-driven on-chip adaptation more challenging.
Recent error-driven online learning methods, such as eligibility propagation (e-prop) algorithm \cite{fre22}, improve learning locality and reduce the temporal and memory overhead associated with conventional backpropagation, but still depend on task-specific learning signals, eligibility-based credit assignment, and surrogate-gradient computation. 
By contrast, STDP performs synaptic updates based solely on local spike-timing correlations, providing a simple and inherently local plasticity mechanism that complements task-oriented error-driven learning.

Implementing such learning in practical neuromorphic systems requires dedicated learning accelerators, yet the hardware implementation of STDP can incur substantial computational and memory-access overhead \cite{liu24}.
A typical SNN may contain tens of thousands of neurons and hundreds of thousands of synapses \cite{li21}, \cite{nei14}, resulting in a large number of local STDP operations for synaptic weight updates. Consequently, neuromorphic processors often require parallel STDP processing modules to sustain the learning throughput of large-scale SNNs. Moreover, event-driven spike communication and synaptic updates lead to irregular and bursty memory accesses, which can lead to contention and data-transfer bottlenecks at high spike rates. These characteristics limit the efficiency of STDP-based learning on conventional von Neumann architectures and have motivated the development of dedicated STDP learning engines and FPGA/ASIC-based neuromorphic accelerators.

In this context, \cite{lam19} presented a digital triplet-based STDP (TSTDP) implementation using low-resolution shift multipliers instead of floating-point multipliers, substantially reducing FPGA resource consumption while preserving biologically plausible plasticity. Other arithmetic-level approximations, such as piece-wise linear (PWL) functions, have also been employed to simplify STDP computation \cite{nou18}. However, these methods mainly reduce arithmetic complexity while leaving the underlying learning procedure and hardware overhead relatively complex. More recently, \cite{zha25} proposed implicit-timing on-chip STDP learning (ImSTDP), which derives spike-timing differences from spike indices and uses a look-up table for weight updates, thereby eliminating counter-based timing computation and improving throughput. Nevertheless, index-difference calculation and table-based updates still incur hardware overhead, while the use of precomputed values reduces flexibility and leaves timing-quantization errors uncompensated. Similarly, \cite{sun22} approximated the STDP curve using a linear function to avoid base-$e$ exponential computation, significantly reducing resource and energy costs. However, without explicit error compensation, the approximation deviates from the original STDP dynamics and degrades biological plasticity.

To address the above limitations, this paper proposes an Intrinsic-Timing Power-of-Two STDP (ITP-STDP) on-chip learning engine that integrates an approximate LIF neuron model with a lightweight local ITP-STDP learning rule, achieving low resource utilization, high energy efficiency, and high operating frequency.
ITP-STDP replaces the base-$e$ exponential function in conventional STDP with a base-$2$ formulation and compensates for the resulting scaling error using a simple $2/3$ approximation. This formulation enables the spike-timing difference to be directly derived from the spike-history shift register, eliminating explicit time-difference computation. Moreover, the spike history and corresponding weight updates share a unified representation, removing the need for LUT-based storage of precomputed updates and reducing the core learning datapath to simple register-based storage, readout, and shift operations. A $4 \times 4$ fully connected prototype is implemented to demonstrate the proposed architecture. Neuronal connectivity is established through direct readout of the spike-history registers, which simplifies the interconnection and data path. The modular structure further enables straightforward scaling to larger SNNs through module replication. The proposed method is systematically evaluated from the synapse level to the network level using a synaptic drift model and SNN experiments, and is further implemented and characterized on both FPGA and ASIC platforms. The main contributions of this paper are summarized as follows:

\begin{enumerate}
    \item An intrinsic-timing power-of-two STDP (ITP-STDP) method is proposed to reduce STDP computational complexity while compensating for approximation errors.
    \item Based on ITP-STDP, a dedicated hardware architecture is developed and evaluated on both FPGA and ASIC platforms against state-of-the-art STDP implementations. The FPGA implementation achieves up to $5.10\times$ higher throughput and $87.77\times$ higher energy efficiency, with up to 99.19\% reduction in logic resources. In 28-nm ASIC implementation, it achieves $5.65\times$--$8.70\times$ higher operating frequency and up to $300.90\times$ higher energy efficiency, while reducing area by up to 98.33\%.
    \item A scalable prototype ITP-STDP learning engine hardware architecture, consisting of four presynaptic neurons fully connected to four postsynaptic neurons, is designed and validated through FPGA and ASIC implementations. Complementary software-based network-level evaluations across datasets of varying complexity, from MNIST to DVS128 Gesture, and different SNN scales demonstrate performance comparable to conventional STDP, with negligible accuracy degradation.
    \item A mean-field synaptic drift model is developed to evaluate the weight-update dynamics of ITP-STDP. The results show that the equilibrium point of the compensated 8-bit ITP-STDP deviates from that of the original STDP by only 3.28\%, corresponding to a 90.0\% reduction in error compared with the uncompensated ITP-STDP. 
\end{enumerate}
The remainder of this paper is organized as follows: Section \ref{chap:2} provides a review of the STDP algorithm, including its weight-update rules and spike-pairing schemes, as well as the LIF neuron model and the related approximate computing algorithm used in the proposed learning engine. Section \ref{chap:3} introduces the proposed ITP-STDP algorithm, together with the architecture of the learning engine built upon it. Section \ref{chap:4} presents the evaluation of ITP-STDP, including the mean-field synaptic drift model, inter-spike-interval-based spike-history depth selection, and network-level experiments. 
Section \ref{chap:5} presents the hardware architecture of the ITP-STDP and ITP-STDP based learning engine. Section \ref{chap:6} reports the FPGA and ASIC implementation results with comparative analysis. Section \ref{chap:7} concludes the paper.


\section{Related Work}
\label{chap:2}
\subsection{Spike-time-dependent plasticity}
Spike-time-dependent plasticity (STDP) is a biologically inspired learning algorithm widely used in spiking neural networks (SNNs) \cite{bai20}. The main function describing its synaptic weight update is expressed as follows \cite{par20}:

\begin{equation}
    \Delta w =
    \begin{cases}
    A_{+}\exp\left(-\Delta t / \tau_{+}\right), &  \Delta t \geq 0 \\
    - A_{-}\exp\left(\Delta t / \tau_{-}\right), &  \Delta t < 0.
    \end{cases}
    \label{eq:stdp}
\end{equation}

In (\ref{eq:stdp}), $\Delta t = t_{post}-t_{pre}$ represents the time difference between the pre-synaptic and post-synaptic spike events, $A_{+}$ and $A_{-}$ denote the scaling factors for synaptic potentiation and depression, respectively, while $\tau_{+}$ and $\tau_{-}$ denote the corresponding time constants that control the decay rate of the weight update. Furthermore, the exponential term in (\ref{eq:stdp}) can be approximated as $2$, yielding \cite{gom18}:
\begin{equation}
    \Delta w =
    \begin{cases}
    A_{+}\cdot2^{-\Delta t / \tau_{+}}, &  \Delta t \geq 0 \\
    - A_{-}\cdot 2^{\Delta t / \tau_{-}}, &  \Delta t < 0.
    \end{cases}
    \label{eq:stdp_app2}
\end{equation}

\subsection{Spike Pairing}
\begin{figure}[ht]
    \centering
    \begin{minipage}[t]{0.8\linewidth}
        \centering
        \includegraphics[width=\textwidth]{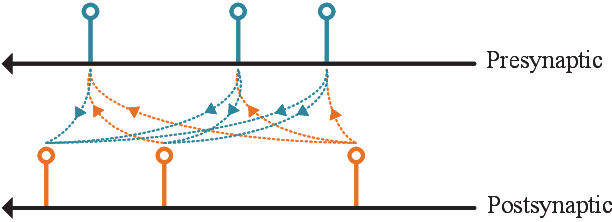}

        \scriptsize{(a) All-to-all spike-pairing.\\ \quad}
    \end{minipage}
    
    \begin{minipage}[t]{0.8\linewidth}
        \centering
        \includegraphics[width=\textwidth]{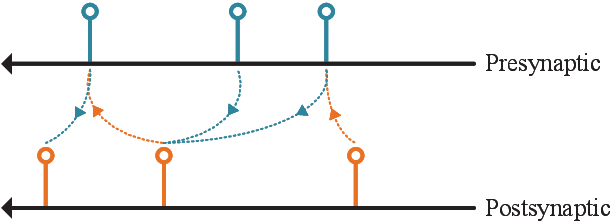}
        
        \scriptsize{(b) Nearest-neighbor pairing.}
    \end{minipage}
    \caption{Spike pairing mechanism.}
    \label{fig:pair}
\end{figure}

There are various spike-pairing mechanisms in STDP. In general, they can be generally be categorized into all-to-all pairing and nearest-neighbor pairing \cite{zho23}. The all-to-all spike-pairing mechanism is illustrated in Fig. \ref{fig:pair}. Its weight variation, $\Delta w$, is obtained by accumulating the contributions of all spike pairs, as given in

\begin{equation}
\Delta w_{all} =
\begin{cases}
\sum_{\Delta t} \mathrm{LTP}(\Delta t), & \Delta t = t_{\mathrm{post}} - t_{\mathrm{pre,all}} \geq 0 \\
\sum_{\Delta t} \mathrm{LTD}(\Delta t), & \Delta t = t_{\mathrm{post,all}} - t_{\mathrm{pre}} < 0,
\end{cases}
\label{eq:allstdp}
\end{equation}
where, $LTP(\Delta t)$ and $LTD(\Delta t)$ represent the potentiation and depression components of the STDP rule, respectively. In contrast, nearest-neighbor pairing considers only the closest Long-Term Depression (LTD) and Long-Term Potentiation (LTP) interactions, as
\begin{equation}
\Delta w_{nn} =
\begin{cases}
\mathrm{LTP}(\Delta t), & \Delta t = t_{\mathrm{post}} - t_{\mathrm{pre,nearest}} \geq 0 \\
\mathrm{LTD}(\Delta t), & \Delta t = t_{\mathrm{post,nearest}} - t_{\mathrm{pre}} < 0.
\end{cases}
\label{eq:nnstdp}
\end{equation}

The proposed Intrinsic-Timing Power-of-Two STDP (ITP-STDP) is capable of supporting both pairing mechanisms. However, the proposed learning engine adopts the nearest-neighbor pairing mechanisms to reduce computational complexity and lower hardware resource utilization. 

\subsection{LIF neuron model}
The discrete-time iteration equation of the membrane potential \(v\) in the LIF neuron model \cite{wu21} is given by
\begin{align}\label{eq.LIF}
    V[t+1] = e^{-\frac{1}{\tau}}\cdot(V[t]-E_{rest}) + E_{rest} + I.
\end{align}
In \eqref{eq.LIF}, \(\tau\) represents the membrane time constant. When \(v\) reaches the threshold, the neuron generates a spike and updates \(v\) to the resting potential as
\begin{align}\label{eq.thre}
    if\,\,V > V_{th},\,\,then \,\,V = E_{rest}. 
\end{align}

\subsection{Logarithmic Linear Segmented Multiply}
The adopted LIF neuron model is constructed based on Logarithmic Linear Segmented Multiply (LLSMu) \cite{x25}. LLSMu integrates the Mitchell approximate method \cite{mit62} with Karatsuba multiplication \cite{kar63}. In this scheme, large-bitwidth multiplication is first decomposed into smaller parallel multiplications using the Karatsuba method, and these smaller multiplications are then carried out using the Mitchell approximate method with logarithmic linear approximation. 

The computation of the LLSMu is formulated as follows. The $2N$-bit operands $A$ and $B$ are first decomposed into their high and low $N$-bit parts as
\begin{equation}
A = 2^N H_A + L_A,\qquad
B = 2^N H_B + L_B .
\end{equation}

In addition, for two operands $x$ and $y$, the mitchell approximate method based multiplication \cite{saa18} is defined as
\begin{equation}
\mathcal{M}(x,y)=
2^{k_x+k_y}
\begin{cases}
c+\dfrac{x}{2^{k_x}}+\dfrac{y}{2^{k_y}}-1,
& \delta < 1, \\[6pt]
\dfrac{c}{2}+\dfrac{x}{2^{k_x}}+\dfrac{y}{2^{k_y}}-2,
& \delta \ge 1,
\end{cases}
\end{equation}
where
\begin{equation}
k_x=\lfloor \log_2 x \rfloor,\qquad
k_y=\lfloor \log_2 y \rfloor,
\end{equation}
\begin{equation}
\delta=\dfrac{x}{2^{k_x}}+\dfrac{y}{2^{k_y}}-2.
\end{equation}
Here $k_x$ and $k_y$ denote the leading-one positions of $x$ and $y$, respectively. The constant $c$ is an error-compensation factor and is set to $0.08333$. Based on the Karatsuba decomposition, the partial products are approximated as
\begin{equation}
m_0=\mathcal{M}(L_A,L_B),\qquad
m_1=\mathcal{M}(H_A,H_B),
\end{equation}
\begin{equation}
m_2=\mathcal{M}(H_A+L_A,H_B+L_B).
\end{equation}
The cross term is then obtained by
\begin{equation}
s_3=m_2-m_0-m_1,
\end{equation}
and the final approximate product is reconstructed as
\begin{equation}
P(A,B)=2^{2N}m_1+2^Ns_3+m_0.
\end{equation}
Equivalently,
\begin{equation}
\begin{aligned}
P(A,B)
={}& 2^{2N}\mathcal{M}(H_A,H_B) \\
&+ 2^N \big[ \mathcal{M}(H_A+L_A,H_B+L_B) - \mathcal{M}(H_A,H_B) \\
&\qquad - \mathcal{M}(L_A,L_B) \big] + \mathcal{M}(L_A,L_B).
\end{aligned}
\end{equation}

\section{Intrinsic-Timing Power-of-Two STDP learning engine}
\label{chap:3}
\subsection{Intrinsic-Timing Power-of-Two STDP (ITP-STDP)}
To obtain a discrete-time form, (\ref{eq:stdp}) is first discretized by setting $\Delta t=k \Delta T$, yielding

\begin{equation}
    \Delta w(k) =
    \begin{cases}
    A_{+}\exp\left(-k\frac{\Delta T}{ \tau_{+}}\right), &  k \geq 0 \\
    - A_{-}\exp\left(k\frac{\Delta T}{ \tau_{-}}\right), &  k < 0.
    \end{cases}
    \label{eq:disstdp}
\end{equation}
Then, by setting $\tau_{+}=\tau_{-}$ and
\begin{equation}
\hat{\Delta T} = k\frac{\Delta T}{\tau_{+}} = k\frac{\Delta T}{\tau_{-}},
\label{eq:time_dif}
\end{equation}
the expression can be further written as
\begin{equation}
    \Delta w(\hat{\Delta T}) =
    \begin{cases}
    A_{+}e^{-\hat{\Delta T}}, & \hat{\Delta T} \geq 0 \\
    - A_{-}e^{\hat{\Delta T}}, & \hat{\Delta T} < 0.
    \end{cases}
    \label{eq:dis_tran_stdp}
\end{equation}

In the discrete-time implementation, the STDP update depends on the normalized timing difference rather than on the physical time constant explicitly. As result, the weight update rule can be directly written as an exponential function of $\Delta \hat{T}$. Furthermore, by setting 
\begin{equation}
    \hat{\Delta T_2} = \frac{\hat{\Delta T}}{\ln 2},
    \label{eq:error}
\end{equation}
the weight-update expression can be further written as follows:
\begin{equation}
    \Delta w(\hat{\Delta T_2}) =
    \begin{cases}
    A_{+} \cdot 2^{-\hat{\Delta T_2}}, &  \hat{\Delta T_2} \geq 0 \\
    - A_{-} \cdot 2^{\hat{\Delta T_2}}, & \hat{\Delta T_2} < 0.
    \end{cases}
    \label{eq:2stdp}
\end{equation}

\begin{figure}[ht]
    \centering
    \begin{minipage}[t]{0.49\linewidth}
        \centering
        \includegraphics[width=1\textwidth]{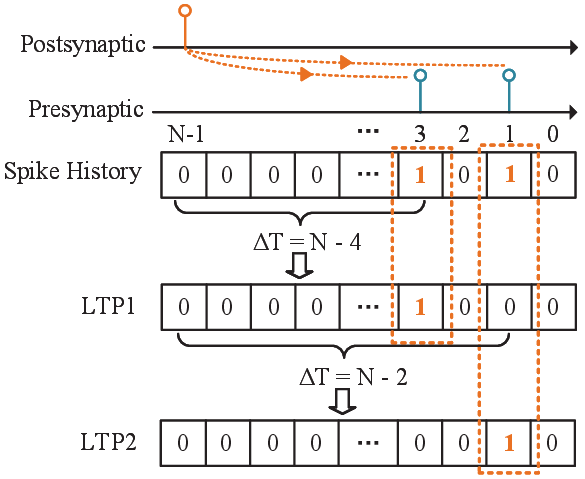}
        
        \scriptsize{(a) LTP.}
    \end{minipage}
    \begin{minipage}[t]{0.49\linewidth}
        \centering
        \includegraphics[width=1\textwidth]{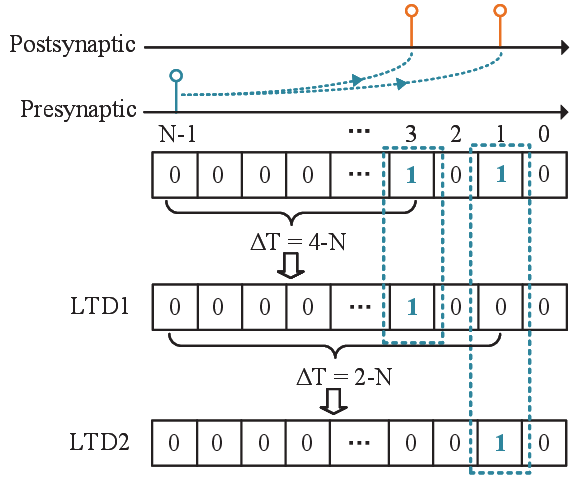}  
        
        \scriptsize{(b) LTD.}
    \end{minipage}
    \caption{LTP and LTD computation mechanism in ITP-STDP.}
    \label{fig:weight}
\end{figure}

In hardware, temporal differences cannot be represented continuously, and the temporal resolution is limited to one clock cycle. As illustrated in Fig. \ref{fig:weight}, spikes are sequentially stored in the spike-history registers according to their corresponding clock cycles, forming a binary bit stream in which each bit indicates the presence or absence of a spike within a specific clock cycle. Therefore, the discretized temporal difference, $\hat{\Delta T}_2$, which is determined by the relative positions of spikes in the spike history, can also be represented as a binary array.
Therefore, by approximating the $e$-based exponential with a base-$2$ exponential, the $\Delta w$ can be directly evaluated from the binary encoding of $\hat{\Delta T_2}$. Consequently, the exponential computation is eliminated and simplified to reading the most significant bit (MSB) of the binary representation, as illustrated in Fig. \ref{fig:weight}. When the nearest-neighbor pairing mechanism is adopted, LTD or LTP can be determined by accessing only the spike at MSB position in the spike history. 

\begin{figure}[ht]
    \centering
    \includegraphics[width=1\linewidth]{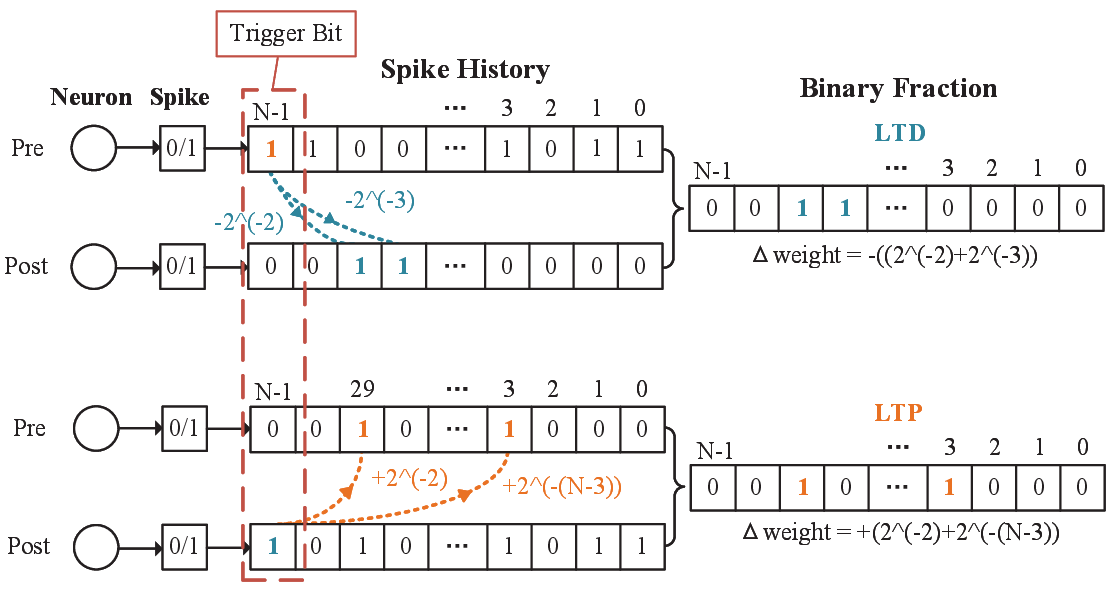}
    \caption{ITP-STDP principle}
    \label{fig:ltpstdp}
\end{figure}

Moreover, in the all-to-all pairing case, this approximation offers further advantages. Specifically, by representing the spike sequence as a binary fixed-point value with a single integer bit, the absolute value of the weight update can be computed directly. As shown in Fig. \ref{fig:weight} and Fig .\ref{fig:ltpstdp}, the the accumulation term in (\ref{eq:allstdp}) for the all-to-all spike-pairing mechanism can be inherently implemented through the composition of the binary-fraction representation. Consequently, in the all-to-all spike-pairing mechanism, ITP-STDP avoids not only the exponential computation itself but also the additional accumulation operation. Moreover, LTD and LTP are triggered separately by the first bit of the corresponding spike sequences.


Since the weight update is directly determined from the time difference, 
the resulting approximation error in the weight change is characterized 
by~\eqref{eq:error}. Therefore, to compensate for the error 
introduced by the approximation, the calculated weight change should be 
divided by $\ln 2$, as follows:

\begin{equation}
     \Delta w(\hat{\Delta T}) = \frac{1}{\ln 2} \cdot \Delta w(\hat{\Delta T_2}).
    \label{eq:error}
\end{equation}

For a hardware-efficient implementation, the compensation factor $1/\ln 2$ 
is approximated by $3/2$. Consequently, the compensation can be realized using 
only a bit-shift operation and an addition, yielding the following expression:

\begin{equation}
    \widetilde{\Delta w(\Delta T)} =
    \begin{cases}
   \frac{3}{2} \cdot A_{+} \cdot 2^{-\Delta T}, &  \Delta T \geq 0 \\
    -  \frac{3}{2} \cdot A_{-} \cdot 2^{\Delta T}, & \Delta T < 0.
    \end{cases}
    \label{eq:2stdp}
\end{equation}

\subsection{Computational Complexity of ITP-STDP}
Since the weight-scaling factors $A_+$ and $A_-$ are typically fixed and chosen close to unity, they are approximated by sums of powers of two in the hardware implementation, enabling multiplication to be replaced by simple shift-and-add operations. Although both conventional STDP and ITP-STDP exhibit $O(1)$ asymptotic complexity per synaptic update, their arithmetic costs differ substantially. Table \ref{tab:arithmetic_comp} summarizes the operations required for a single weight update in the two methods.

\begin{table}[ht]
\centering
\caption{Arithmetic Opeartion Comparison Between STDP and ITP-STDP}
\label{tab:arithmetic_comp}
\begin{tabular}{cccc}
\toprule \toprule
Operation & STDP & ITP-STDP w/o Comp. & ITP-STDP \\
\midrule
EXP        & 1  & 0     & 0 \\
MUL        & 1  & 0     & 0 \\
ADD        & 1  & 1     & 2 \\
SHIFT      & 0  & 1     & 2 \\
\midrule
Transcendental op. & 1 & 0 & 0\\
\bottomrule
\bottomrule
\end{tabular}
\end{table}

Conventional STDP requires one multiplication, one addition, and one base-$e$ exponential evaluation to compute the synaptic weight change, where the exponential evaluation constitutes a transcendental operation. In ITP-STDP, the multiplication is approximated by  shift-and-add operations, while the addition associated with explicit time-difference computation is eliminated by directly obtaining the temporal difference from the spike-history representation. When compensation is applied, the $3/2$ scaling factor is implemented using one shift and one addition instead of a multiplication. Thus, the weight-change computation in the uncompensated and compensated ITP-STDP requires only one shift-and-add operation and two shift-and-add operations, respectively, without any transcendental computation.

\subsection{Structure of ITP-STDP Learning Engine}
\begin{figure}[ht]
    \centering
    \includegraphics[width=\linewidth]{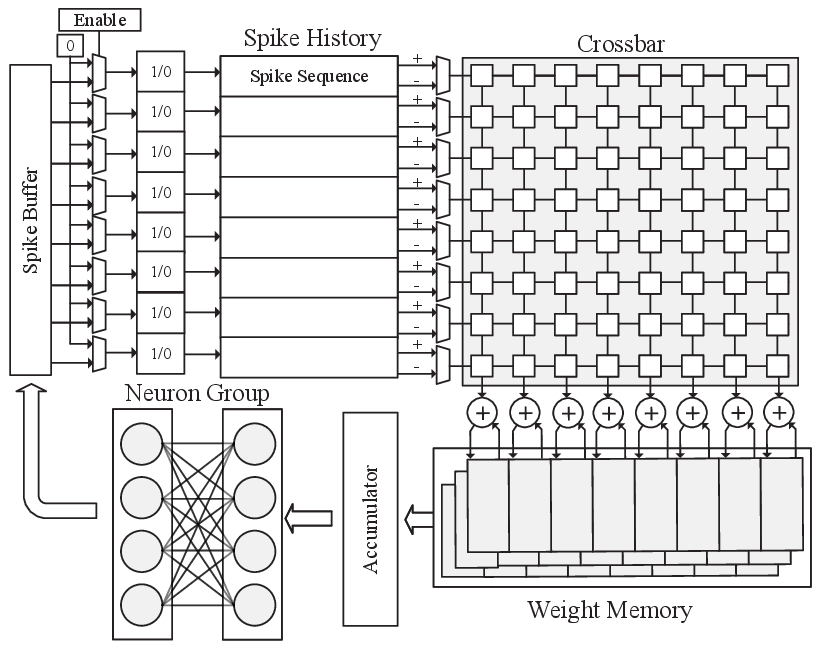}
    \caption{Structure of ITP-STDP Learning Engine}
    \label{fig:structure}
\end{figure}

The ITP-STDP learning engine is designed to operate almost entirely through direct read and write access to spike sequences. Such operations enable both the computation of the weight update $\Delta w$ and the direct establishment of inter-neuron connections through spike-sequence reading. As illustrated in Fig. \ref{fig:ltpstdp}, the proposed ITP-STDP architecture uses the first bit of different spike histories as a trigger to determine the sign of the weight update and whether the associated neurons should be connected. Moreover, as shown in Fig. \ref{fig:structure}, the proposed architecture employing the ITP-STDP learning engine is applicable to neural topologies with arbitrary connectivity. The Spike History module stores the spike sequences of all neurons, which also encode the corresponding LTD and LTP information. A crossbar is employed to select the LTD and LTP signals of arbitrary neurons, enabling the establishment of connections between any neuron pair. Subsequently, $\Delta w$ is then directly applied to update the synaptic weights, which are stored in the weight memory and provided back to the neurons. Therefore, the ITP-STDP learning engine requires only a weight-update adder as its arithmetic unit, while all other components are devoted to weight and spike storage. Since inter-neuron connectivity is also realized through storage access, the proposed learning engine exhibits very low hardware resource utilization and power consumption.

Moreover, the proposed architecture is inherently scalable. Although only a simple $4 \times 4$ fully connected prototype is implemented in Fig. \ref{fig:structure}, the architecture can be readily extended to support larger-scale SNNs by replicating the proposed processing structure.

\section{Evaluation of proposed design}
\label{chap:4}
To further validate and evaluate the proposed ITP-STDP, an emulator for both STDP and ITP-STDP was developed based on PyTorch and SpikingJelly.

\subsection{Inter-Spike Interval-Based Spike-History Depth Selection}

With the proposed computational optimizations, the main hardware overhead in the ITP-STDP design is reduced to the shift-register array used for spike-history storage. The depth of the stored spike history determines the temporal window over which ITP-STDP can capture spike timing information. To further reduce hardware resource consumption while retaining sufficient spike-timing information, an inter-spike interval (ISI)-based temporal-window analysis is introduced. To determine an appropriate spike-history depth, the ISI distributions of spike sequences transformed from three different datasets were statistically analyzed. 
The analyzed data comprised two image datasets, MNIST and Fashion-MNIST, an event-based DVS128 Gesture dataset capturing asynchronous spatiotemporal visual events, and a private time-series motor fault diagnosis dataset containing current and magnetic flux variation signals \cite{xi25, xia26}.

The input data are represented as a sequence $x=\{x_i\}_{i=1}^{N}$, where $x_i$ denotes an input element, such as a pixel intensity or a time-series sample. Each $x_i$ is then normalized using min-max normalization as follows:

\begin{equation}
x_i^{\mathrm{norm}} = \frac{x_i - x_{\min}}{x_{\max} - x_{\min}}.
\end{equation}
where $x_{\min}$ and $x_{\max}$ denote the minimum and maximum values within the corresponding input sample, respectively. After normalization, the input value $x_i^{\mathrm{norm}} \in [0,1]$ is encoded into a binary spike sequence using rate coding \cite{wan23}. At each encoding time step $t$, a random variable $r_i(t)$ is drawn from the uniform distribution $\mathcal{U}(0,1)$, and the spike event is generated as

\begin{equation}
s_i(t)=
\begin{cases}
1, & r_i(t)<x_i^{\mathrm{norm}},\\
0, & r_i(t)\geq x_i^{\mathrm{norm}}.
\end{cases}
\end{equation}
This process is equivalent to Bernoulli sampling, where the spike 
probability is given by $P(s_i(t)=1)=x_i^{\mathrm{norm}}$. Therefore, 
the expected normalized firing rate over $T_{\mathrm{rate}}$ encoding 
steps satisfies

\begin{equation}
\mathbb{E}\left[
\frac{1}{T_{\mathrm{rate}}}
\sum_{t=1}^{T_{\mathrm{rate}}}s_i(t)
\right]
=x_i^{\mathrm{norm}}.
\end{equation}

\begin{figure}[ht]
    \centering
    \begin{minipage}[t]{0.8\linewidth}
        \centering
        \includegraphics[width=\textwidth]{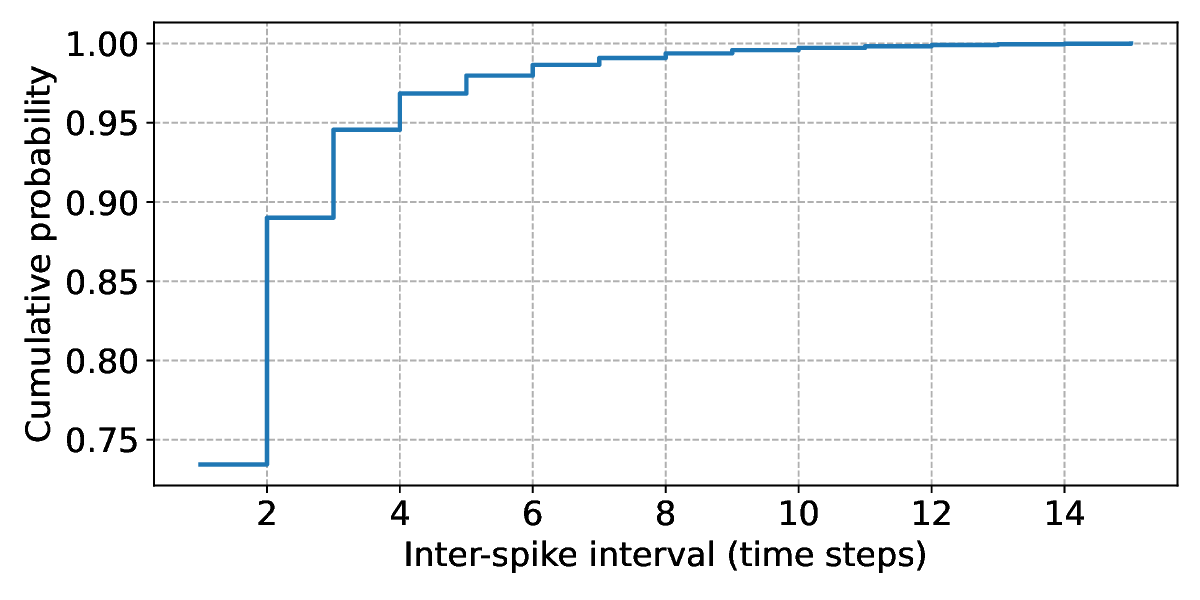}

        \scriptsize{(a)Cumulative Distribution of Inter-Spike Intervals.\\ \quad}
    \end{minipage}  
    \begin{minipage}[t]{0.8\linewidth}
        \centering
        \includegraphics[width=\textwidth]{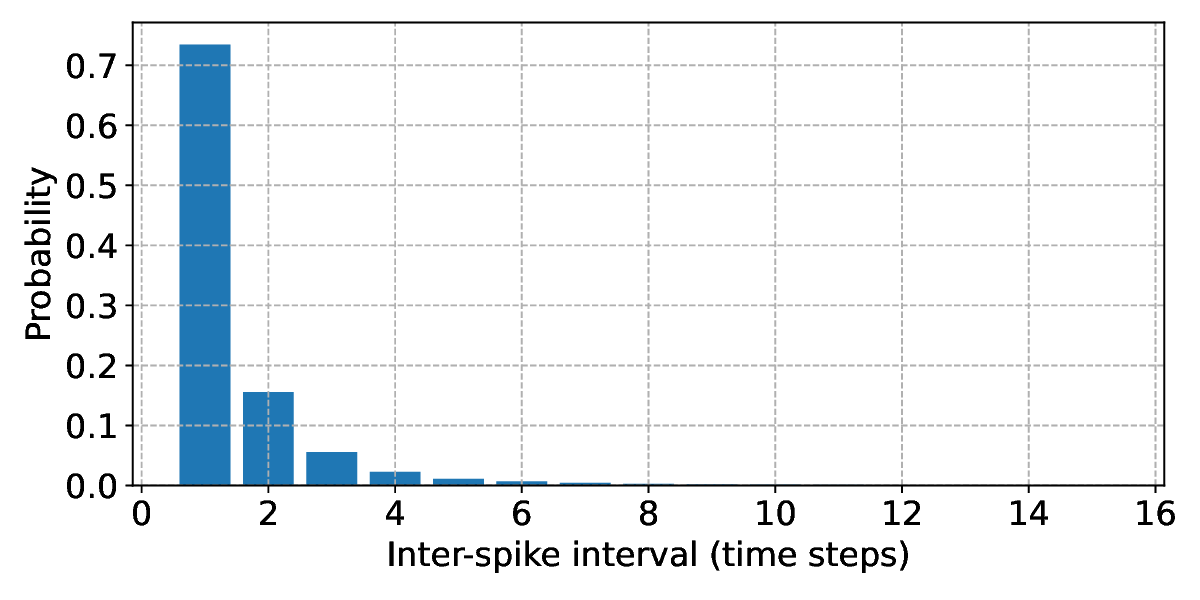}
        
        \scriptsize{(b) Histogram of Inter-Spike Intervals.}
    \end{minipage}
    \caption{ISI-based spike-history depth analysis for representative datasets.}
    \label{fig:isi}
\end{figure}

ISI distribution shown in Fig. \ref{fig:isi}(b) was obtained from 169,754,040 spikes generated by rate coding 10,000 samples from each of the four datasets. It can be observed from the figure that most ISIs fall within 1 to 6 time steps. In comparison, the occurrences at 7 and 8 time steps are negligible, representing only 0.44\% and 0.29\%, respectively. As shown in Fig. \ref{fig:isi}((a), a spike-history depth of 7 covers more than 99\% of the ISI distribution, specifically 99.01\%. Based on this observation, the spike-history depth of ITP-STDP is chosen as 7, and the corresponding weight-update representation adopts an 8-bit width, including one sign bit. In addition, when implemented on FPGA, the ITP-STDP method also supports flexible adjustment of the spike-history depth according to application requirements.

\subsection{Error Analyses of ITP-STDP learning curve}
\begin{figure}[ht]
    \begin{minipage}[t]{0.49\linewidth}
        \centering
        \includegraphics[width=1\textwidth]{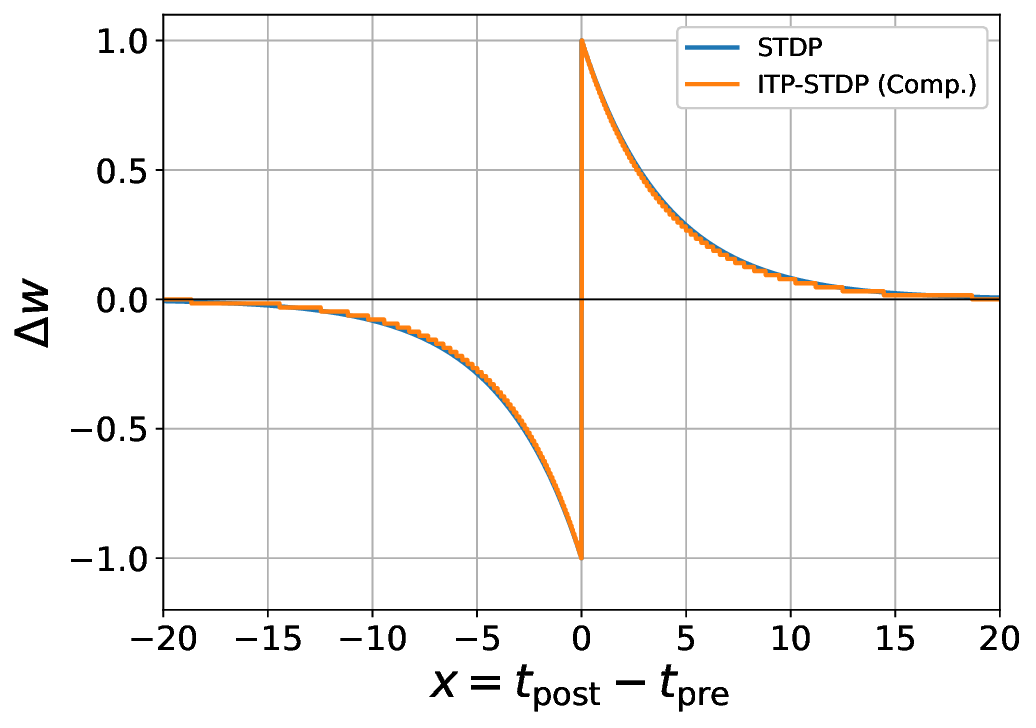}
        
        \scriptsize{(a) ITP-STDP with compensation.}
    \end{minipage}
    \begin{minipage}[t]{0.49\linewidth}
        \centering
        \includegraphics[width=1\textwidth]{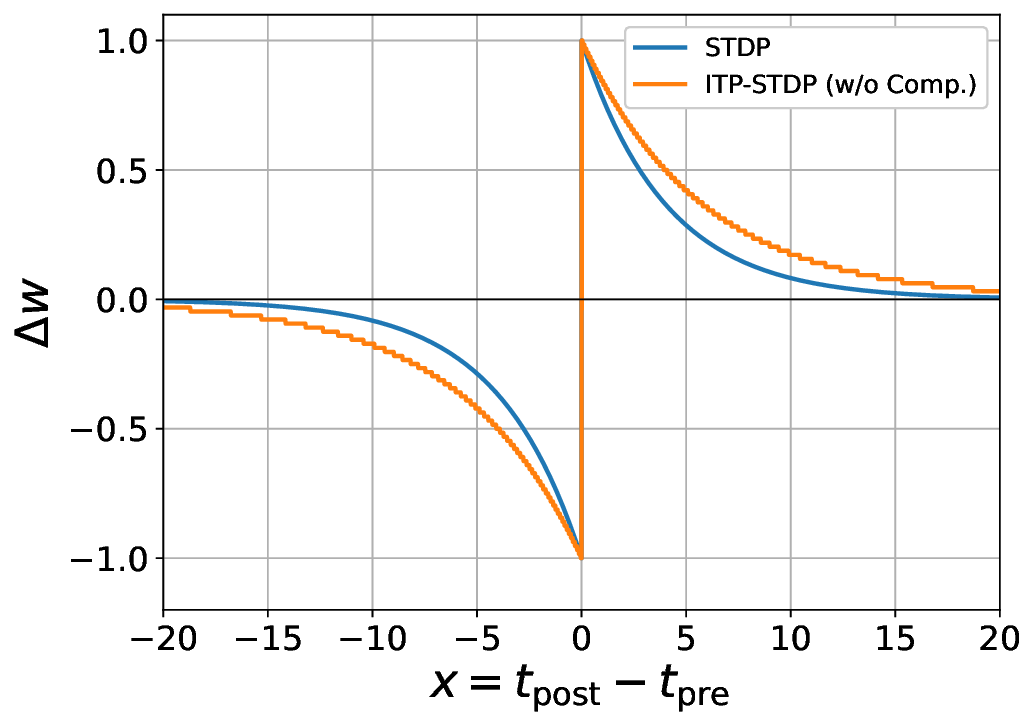}  
        
        \scriptsize{(b) ITP-STDP without compensation.}
    \end{minipage}
    \caption{Comparison of STDP learning curve.}
    \label{fig:stdp_window}
    
\end{figure}

The approximation employed in ITP-STDP, together with 8-bit computation, introduces deviations in the shape of the STDP learning curve as shown in Fig. \ref{fig:stdp_window}. To quantitatively evaluate the similarity between the learning curves of ITP-STDP and the original STDP, the Normalized Root Mean Square Deviation (NRMSD) is adopted. The NRMSD is defined as follows:
\begin{equation}
\begin{aligned}
	&RMSD=\sqrt{\frac{\sum_{x=0}^{n}{(\Delta w_{o}\left(x\right)-\Delta w_{i}\left(x\right))}^2}{n}}\\
    &NRMSD = \frac{RMSD}{max(\Delta w_{o})-min(\Delta w_{o})},
\end{aligned}
\end{equation}
where $\Delta w_o$ and $\Delta w_i$ denote the weight changes obtained using 
the original STDP and ITP-STDP, respectively.

The Python-based emulator shows that the compensated 8-bit ITP-STDP achieves 
an NRMSD of only 0.49\%, compared with 4.45\% for the uncompensated ITP-STDP. 
The proposed output compensation therefore reduces the approximation error by 
88.99\%. The resulting extremely low NRMSD indicates that the compensated 
8-bit ITP-STDP closely matches the original 32-bit floating-point STDP in terms 
of weight-update computation, with only negligible deviation.

\subsection{Dynamical Analyses of ITP-STDP}
To validate the dynamical model of STDP weight updates, a mean-field synaptic drift model was constructed for weight-evolution comparison \cite{bur04}. The normalized update rule for a single synaptic weight can be expressed as follows:
\begin{equation}
    w_{t+1} = \Pi_{[0,1]} \bigl( w_t + \eta\, g(w_t) \bigr).
\end{equation}
In this expression, $w_t$ denotes the synaptic weight at iteration $t$, $\eta$ is the learning rate, and $g(w)$ represents the expected mean weight increment at the current weight value, which is given by
\begin{equation}
    g(w) = \mathbb{E}[\Delta w \mid w] = \int F(x)\, p(x \mid w)\, dx.
    \label{eq:gw}
\end{equation}
In (\ref{eq:gw}), $F(x)$ indicates the weight-update function, corresponding to (\ref{eq:dis_tran_stdp}) for the original STDP and to (\ref{eq:2stdp}) for ITP-STDP, and $p(x \mid w)$ describes the probability distribution of the spike timing difference. A mixture model of background noise and a weight-dependent causal peak is adopted, which can be written as
\begin{equation}
    p(x \mid w) = \bigl(1 - \rho(w)\bigr)\, p_{\mathrm{bg}}(x) + \rho(w)\, p_{\mathrm{c}}(x).
\end{equation}
Here, $p_{\mathrm{bg}}(x)$ corresponds to uncorrelated background spike pairs, while $p_{\mathrm{c}}(x)$ describes causal pre-before-post events concentrated at small positive delays, and $\rho(w)$ represents the weight-dependent mixing coefficient.
This modeling choice captures the idea that stronger synapses are more likely to evoke causal post-synaptic firing, thereby shifting the spike-timing distribution toward the potentiation region. 
To facilitate the analysis, the background noise distribution is modeled by a symmetric Laplace distribution
\begin{equation}
p_{\mathrm{bg}}(x)=\frac{1}{2b}\exp\!\left(-\frac{|x|}{b}\right),
\end{equation}
while the causal distribution is modeled as an exponential distribution: 
\begin{equation}
p_{\mathrm{c}}(x\mid w)=
\begin{cases}
\dfrac{1}{a(w)}\exp\!\left(-\dfrac{x-\mu(w)}{a(w)}\right), & x\ge \mu(w)\\
0, & x<\mu(w),
\end{cases}
\end{equation}
where
\begin{equation}
\mu(w)=m_0+m_1 w,\qquad a(w)=a_0+a_1 w.
\end{equation}
The mixing coefficient is chosen as a saturating function of the synaptic weight,
\begin{equation}
\rho(w)=\frac{\alpha w}{1+\beta w},
\end{equation}
so that the contribution of causal events increases with $w$ but does not grow indefinitely. 

\begin{table}[t]
\centering
\caption{Parameter settings used in the drift analysis.}
\label{tab:parameter_settings}
\begin{tabular}{llll}
\toprule \toprule
Parameter & Value & Value (MNIST) & Description \\
\midrule
$b$        & 5.8  & 28.69     & Background distribution scale \\
$\alpha$   & 0.58 & 0.26    & Mixing coefficient parameter \\
$\beta$    & 4.2  & 19.89     & Mixing coefficient parameter \\
$m_0$      & 0.0  & 4.69     & Base causal delay \\
$m_1$      & 4.5  & 0.13     & Weight-dependent causal delay \\
$a_0$      & 0.5  & 11.05    & Base causal scale \\
$a_1$      & 4.0  & 4.89     & Weight-dependent causal scale \\
$A_{+}$    & 1.0  & 1.0     & Potentiation amplitude \\
$A_{-}$    & 1.125 & 1.125    & Depression amplitude \\
$\tau_{+}$ & 4.0  & 4.0     & Time constant for LTP \\
$\tau_{-}$ & 4.0  & 4.0     & Time constant for LTD \\
$\eta$     & 0.2  & 0.2     & Learning rate \\
    \bottomrule
    \bottomrule
\end{tabular}
\end{table}

\begin{figure*}[ht]
    \centering
    \begin{minipage}[t]{0.24\linewidth}
        \centering
        \includegraphics[width=1\textwidth]{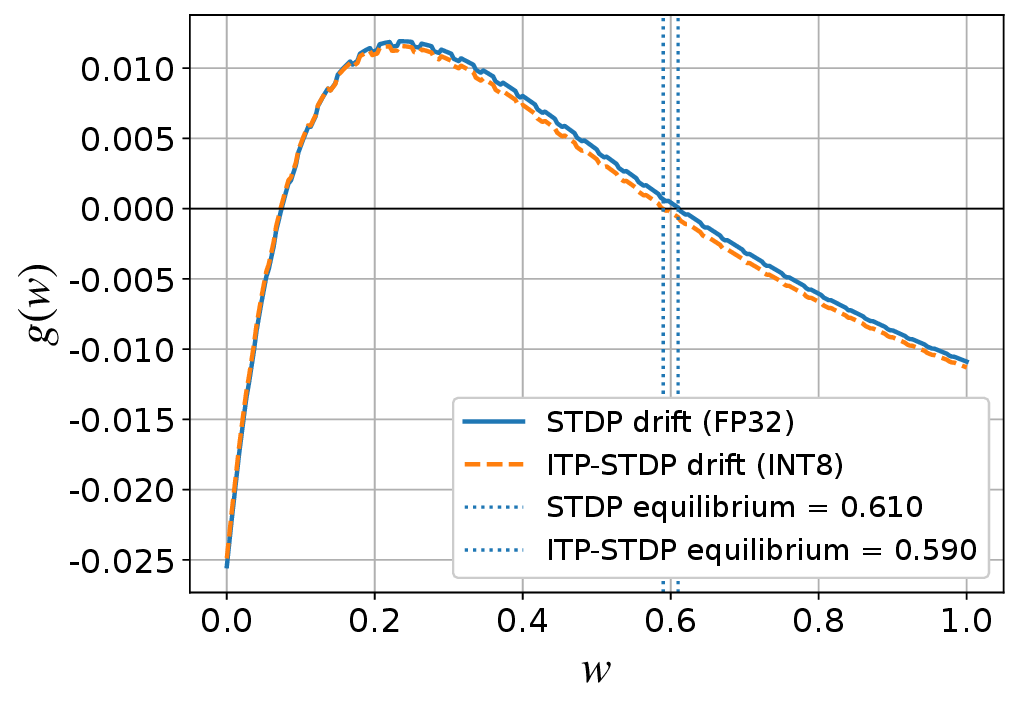}
        
        \scriptsize{(a) Mean Drift Under The Timing Model.}
    \end{minipage}
    \begin{minipage}[t]{0.24\linewidth}
        \centering
        \includegraphics[width=1\textwidth]{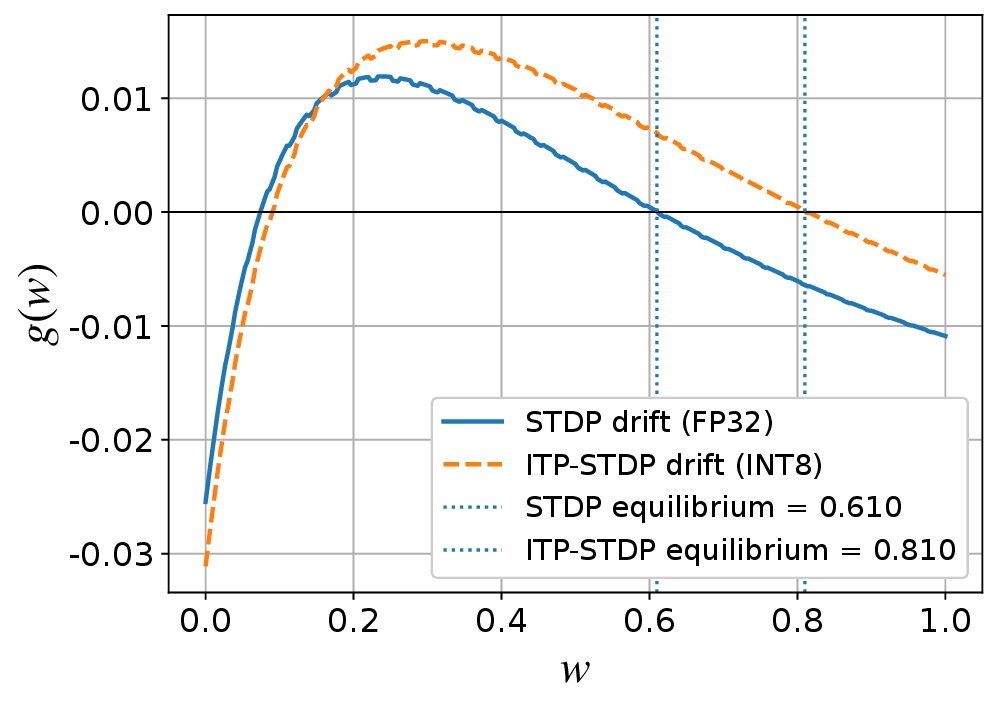}  
        
        \scriptsize{(b) Mean Drift Under The Timing Model.}
    \end{minipage}
    \begin{minipage}[t]{0.24\linewidth}
        \centering
        \includegraphics[width=1\textwidth]{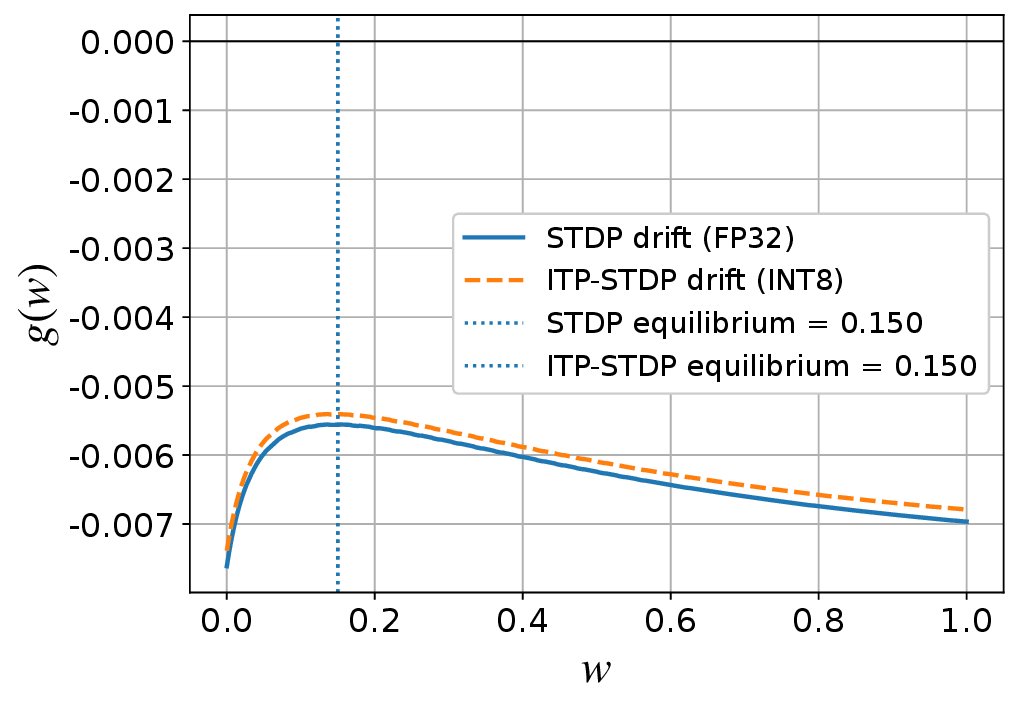}
        
        \scriptsize{(c) Mean Drift Under The Timing Model.}
    \end{minipage}
    \begin{minipage}[t]{0.24\linewidth}
        \centering
        \includegraphics[width=1\textwidth]{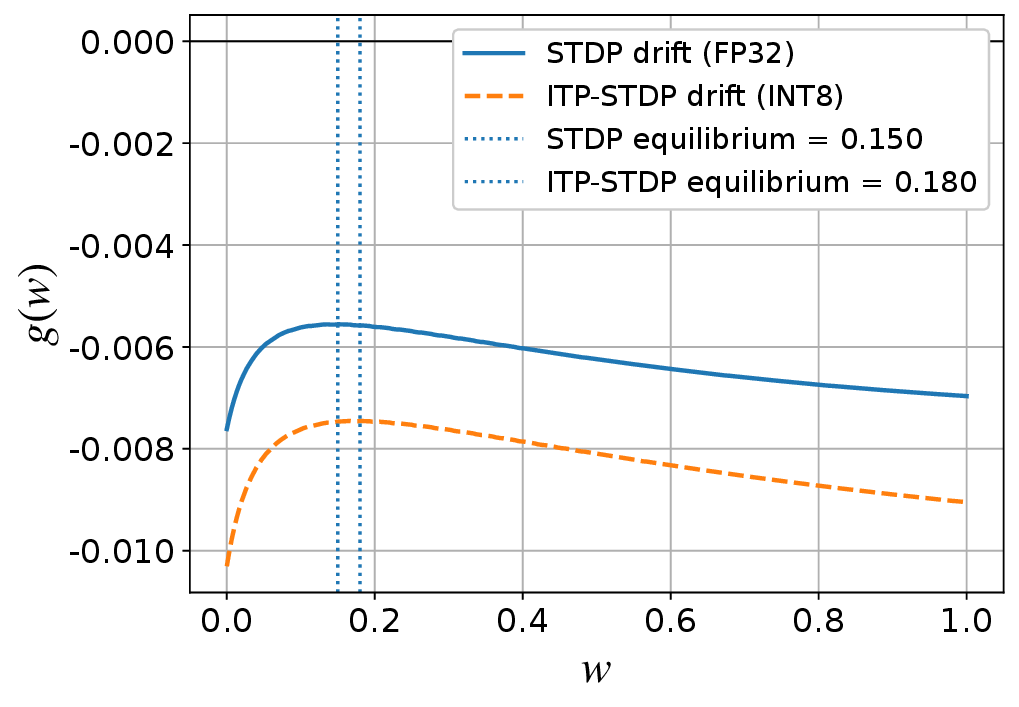}  
        
        \scriptsize{(d) Mean Drift Under The Timing Model.}
    \end{minipage}
    \begin{minipage}[t]{0.24\linewidth}
        \centering
        \includegraphics[width=1\textwidth]{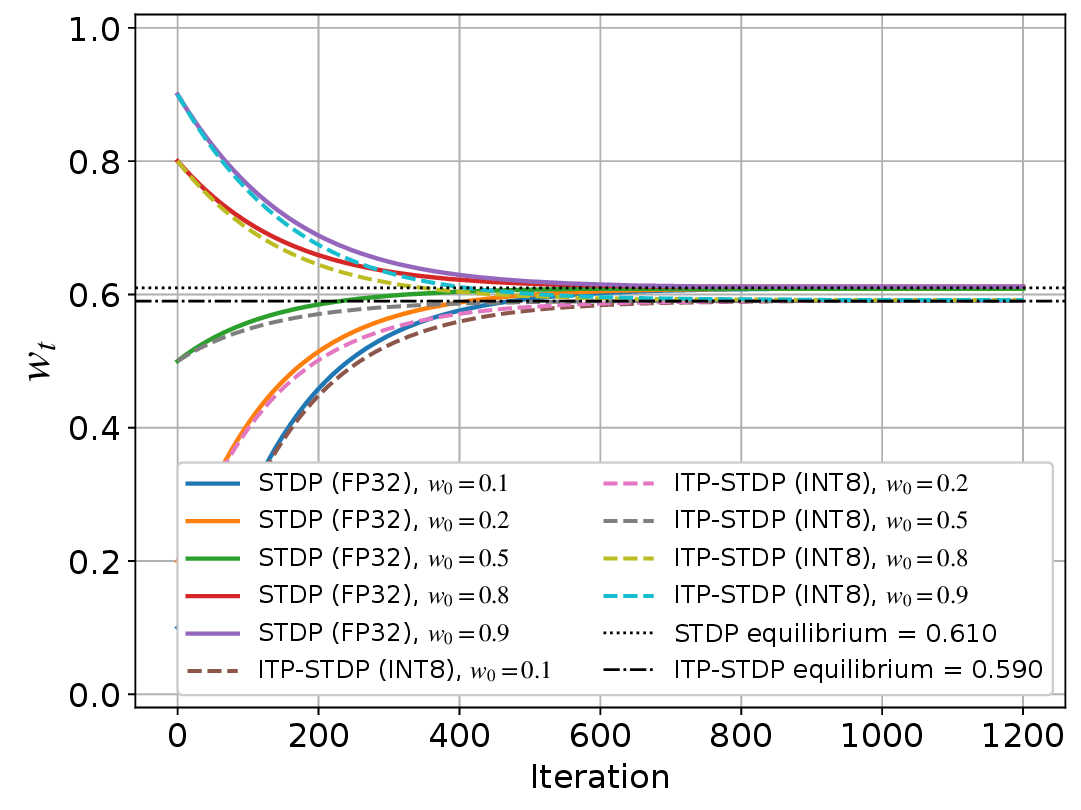}
        
        \scriptsize{(e) Weight Evolution Comparison.}
    \end{minipage}
    \begin{minipage}[t]{0.24\linewidth}
        \centering
        \includegraphics[width=1\textwidth]{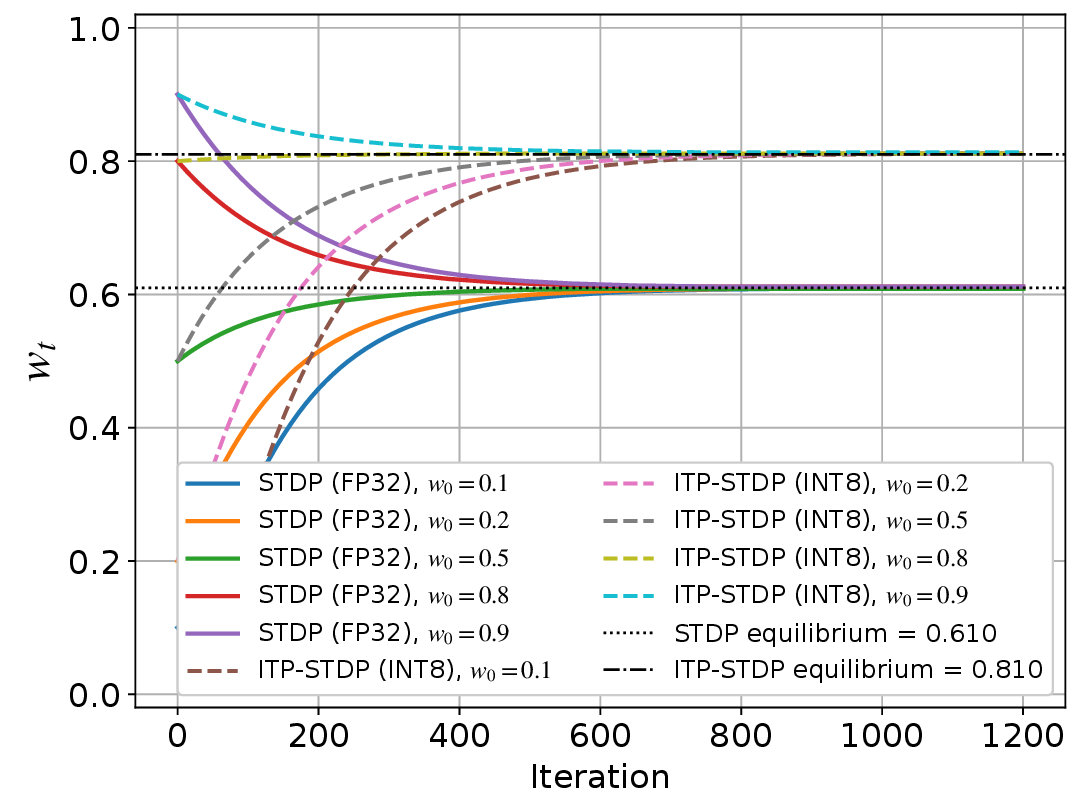}  
        
        \scriptsize{(g) Weight Evolution Comparison.}
    \end{minipage}
    \begin{minipage}[t]{0.24\linewidth}
        \centering
        \includegraphics[width=1\textwidth]{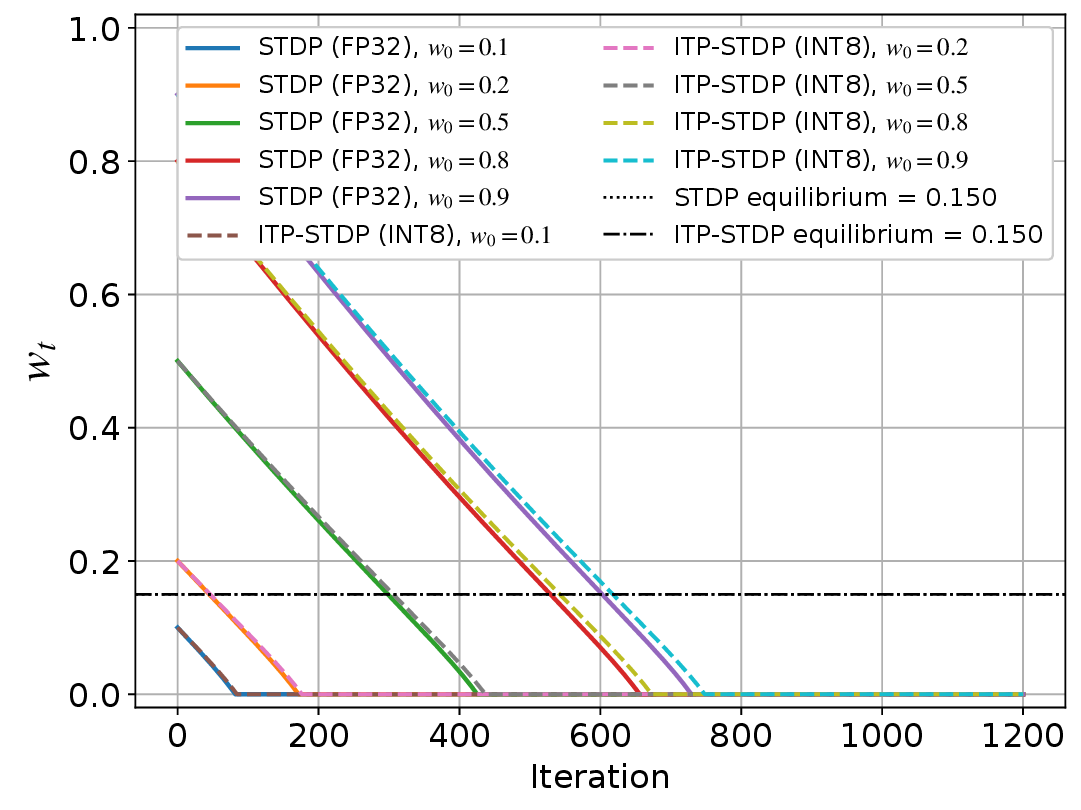}
        \scriptsize{(g) Weight Evolution Comparison.}
    \end{minipage}
    \begin{minipage}[t]{0.24\linewidth}
        \centering
        \includegraphics[width=1\textwidth]{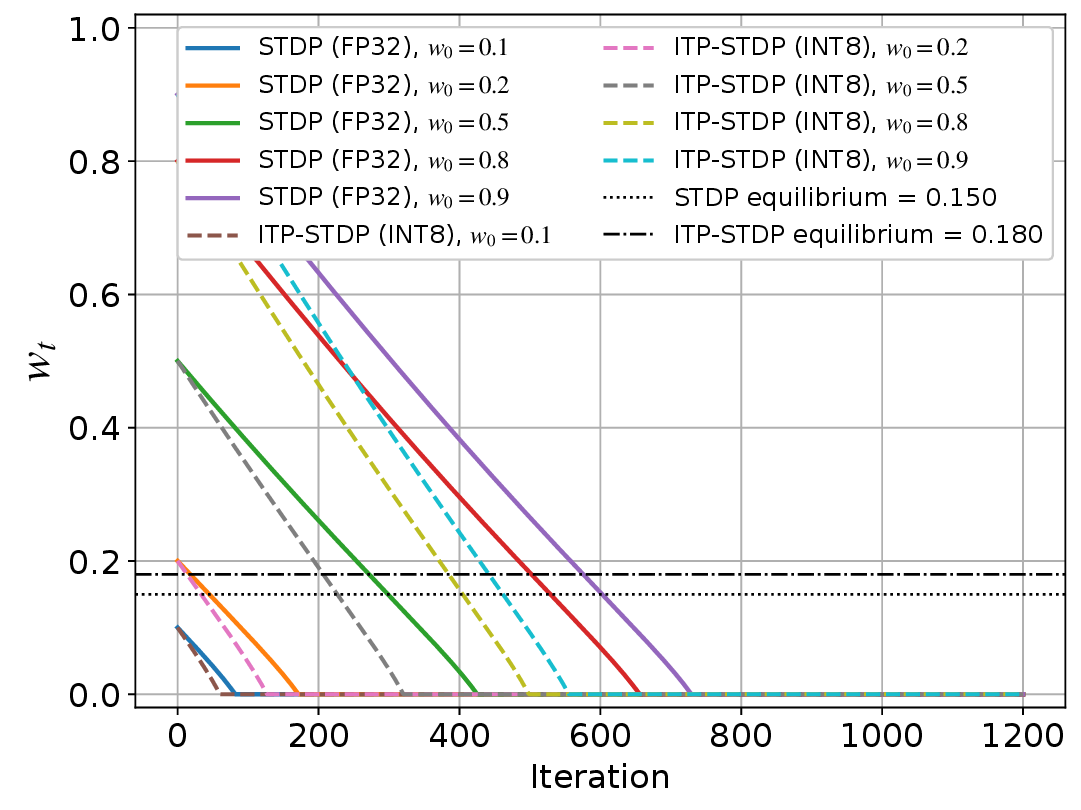}  
        
        \scriptsize{(h) Weight Evolution Comparison.}
    \end{minipage}

    \caption{Results of the Dynamical Analysis of ITP-STDP. The first column corresponds to ITP-STDP with $\tau=4 \times \ln2$, whereas the second column corresponds to ITP-STDP with $\tau=4$.}
    \label{fig:dynamical}
\end{figure*}


Based on this model, the dynamical behaviors of STDP and ITP-STDP are evaluated using two parameter sets listed in Table \ref{tab:parameter_settings} to examine the consistency of the observed dynamics under different parameter conditions. The first is selected to represent an idealized learning-curve shape, while the second is derived by fitting the model to the spike-timing statistics extracted from SNN training on MNIST. The corresponding comparison results are shown in Fig. \ref{fig:dynamical}. Panels (a) and (b) illustrate the local synaptic-weight dynamics under the first parameter set, whereas panels (c) and (d) show those under the second parameter set. Panels (e) and (f) present the corresponding global weight trajectories for the first parameter set, while panels (g) and (h) show those for the second parameter set.

The comparison shows that, after compensation, ITP-STDP exhibits dynamical 
behavior that closely matches that of the original STDP. Under the 
MNIST-fitted parameter set, the local stable points of the synaptic weights 
are identical, while under the second parameter set, they differ by only 
3.28\%. Moreover, the two methods exhibit nearly identical global 
weight-convergence trajectories. These results demonstrate that the 
compensated ITP-STDP preserves the dynamical characteristics and 
weight-evolution behavior of the original STDP, indicating that the 
approximation error introduces only a negligible effect on the synaptic 
weight-update process.


In contrast, without compensation, the approximation in ITP-STDP leads to substantial deviations from the dynamical behavior of the original STDP. The equilibrium points exhibit errors of 32.79\% and 20.00\% under the two parameter sets, respectively, resulting in significantly different weight-convergence curves. These deviations arise from both the use of 8-bit fixed-point arithmetic instead of 32-bit floating-point computation and the direct approximation of the base-(e) exponential function by a base-2 exponential function. These results demonstrate the importance of error compensation in maintaining the original STDP dynamics.


\subsection{ITP-STDP based Spiking Neural Networks}


To further evaluate the impact of the approximation error in ITP-STDP, the use of 8-bit synaptic weights, and the approximate LIF neuron implementation employed in the ITP-STDP learning engine on larger-scale SNNs, a spiking neural network validation framework was developed to assess their effects on the overall performance of the neuromorphic system. The framework adopts a hybrid learning strategy that combines STDP with gradient descent to achieve high classification accuracy.
Within this framework, three different SNN architectures based on the LIF neuron model were trained and evaluated on four datasets, including the relatively simple image-based MNIST dataset, the more challenging Fashion-MNIST dataset, the event-based DVS128 Gesture dataset containing real-world neuromorphic event streams \cite{ami17}, and a motor fault diagnosis dataset collected from laboratory experiments \cite{xi25}.

\begin{figure}[ht]
    \centering
    \includegraphics[width=0.8\linewidth]{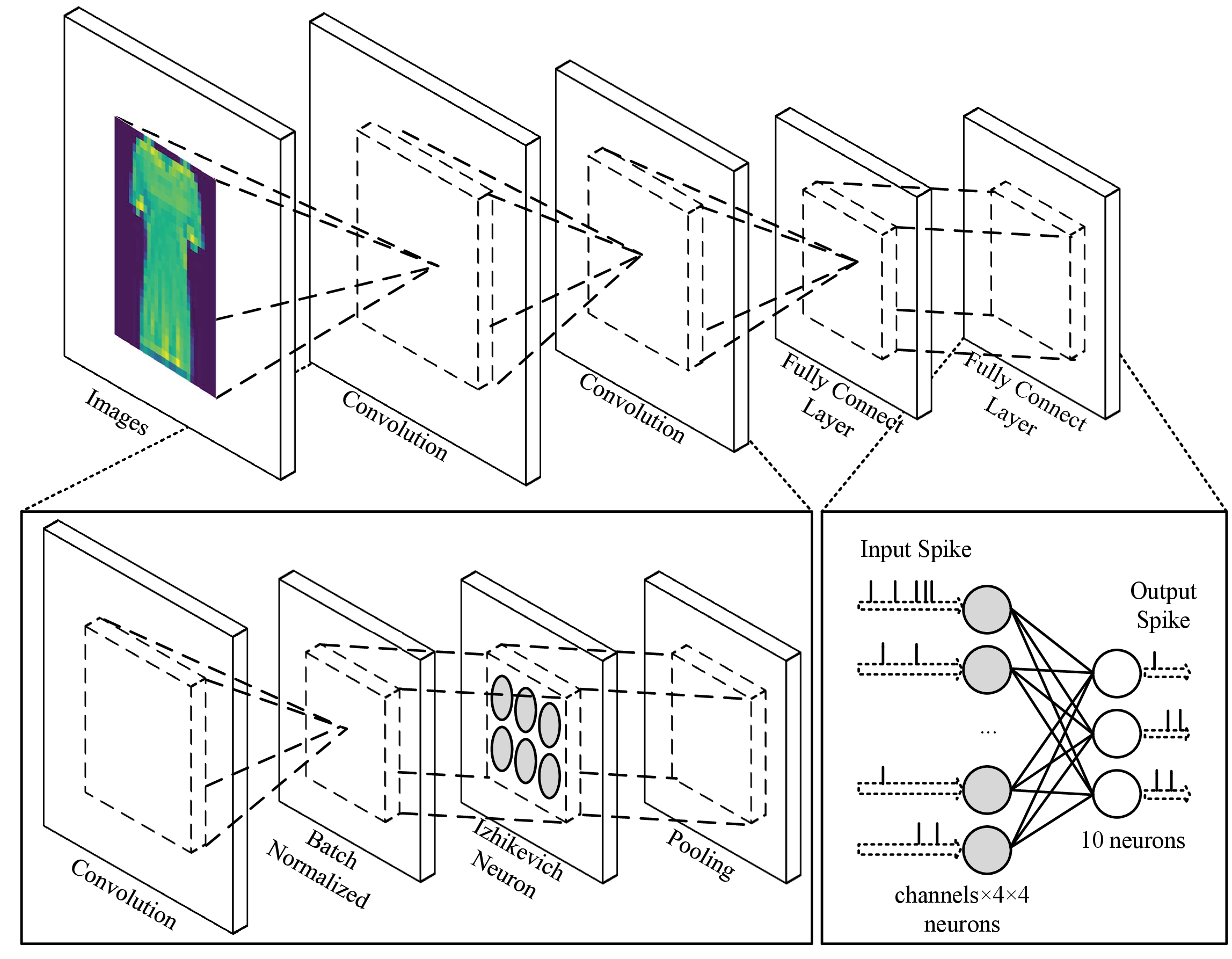}
    \caption{Structure of 6-layers DCSNN.}
    \label{fig:fmnist}
\end{figure}

\begin{figure}[ht]
    \centering
    \includegraphics[width=1\linewidth]{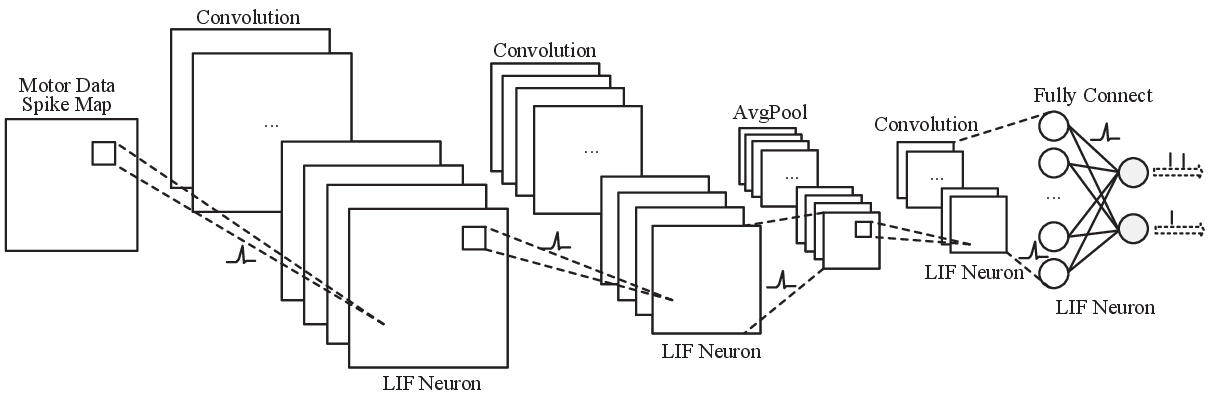}
    \caption{Structure of 5-layers CSNN.}
    \label{fig:motor}
\end{figure}

For the MNIST dataset, a highly lightweight two-layer SNN was adopted, with a fully connected architecture consistent with that shown in Fig. \ref{fig:fmnist}. For the Fashion-MNIST dataset and DVS128 Gesture dataset, a six-layer deep convolutional spiking neural network (DCSNN) was employed to evaluate the performance of a more complex neuromorphic system, as illustrated in Fig. \ref{fig:fmnist}. In addition, for the time-series motor fault diagnosis data, a five-layer convolutional spiking neural network (CSNN) was constructed to address a practical real-world application, as shown in Fig. \ref{fig:motor}.


\begin{table}
    \centering
    \caption{Evaluation results for different STDP methods across network architectures and datasets}
    \resizebox{0.5\textwidth}{!}{
    \renewcommand{\arraystretch}{1.2} 
    \setlength{\tabcolsep}{5pt}     
    \begin{tabular}{ccccc}
        \toprule
        \toprule
        \multirow{2}{*}{\makecell{\\Network}}& \multirow{2}{*}{\makecell{\\Dataset}} &  \multicolumn{3}{c}{STDP} \\
        \cline{3-5}
         &   & Original & \makecell{\rule{0pt}{1.1em}ITP-STDP\\(w/o Comp.)} & \makecell{\rule{0pt}{1.1em}ITP-STDP\\(Comp.)} \\
        \hline
        2-layer SNN  & MNIST & $98.15 \pm  0.03$ & $97.97 \pm 0.06$ & $98.18 \pm 0.05$\\
        6-layer DCSNN  & F-MNIST & $91.19 \pm  0.07$ & $91.12 \pm 0.04$ & $91.08 \pm 0.04$\\
        6-layer DCSNN  & DVS128 & $81.71 \pm  1.12$ & $81.83 \pm 1.22$ & $81.48 \pm 0.40$\\
        5-layer CSNN & Rotor Fault & $90.13 \pm  0.19$ & $89.42 \pm 0.40$ & $90.30 \pm 0.20$\\
        \bottomrule
        \bottomrule
    \end{tabular}}
    \label{tab:nn_comp}
    
    \raggedright \scriptsize{Comp. denotes ITP-STDP with compensation; w/o Comp. denotes ITP-STDP without compensation.}
\end{table}

Table \ref{tab:nn_comp} summarizes the results obtained across multiple random seeds, with all experiments conducted under identical hyperparameter settings. The results show that the ITP-STDP learning engine, with or without compensation, achieves test accuracy comparable to that of the reference system using FP32 STDP and LIF neuron computations. This indicates that the small errors introduced by the approximate computations have only a limited impact on the performance of larger-scale SNNs. Moreover, the ITP-STDP learning engine demonstrates consistently stable performance across different network scales and task complexities.

\section{Hardware Deign and Implementation of proposed ITP-STDP learning engine}
\label{chap:5}
\subsection{Hardware Design of ITP-STDP}

\begin{figure*}[ht]
    \centering
    \includegraphics[width=\linewidth]{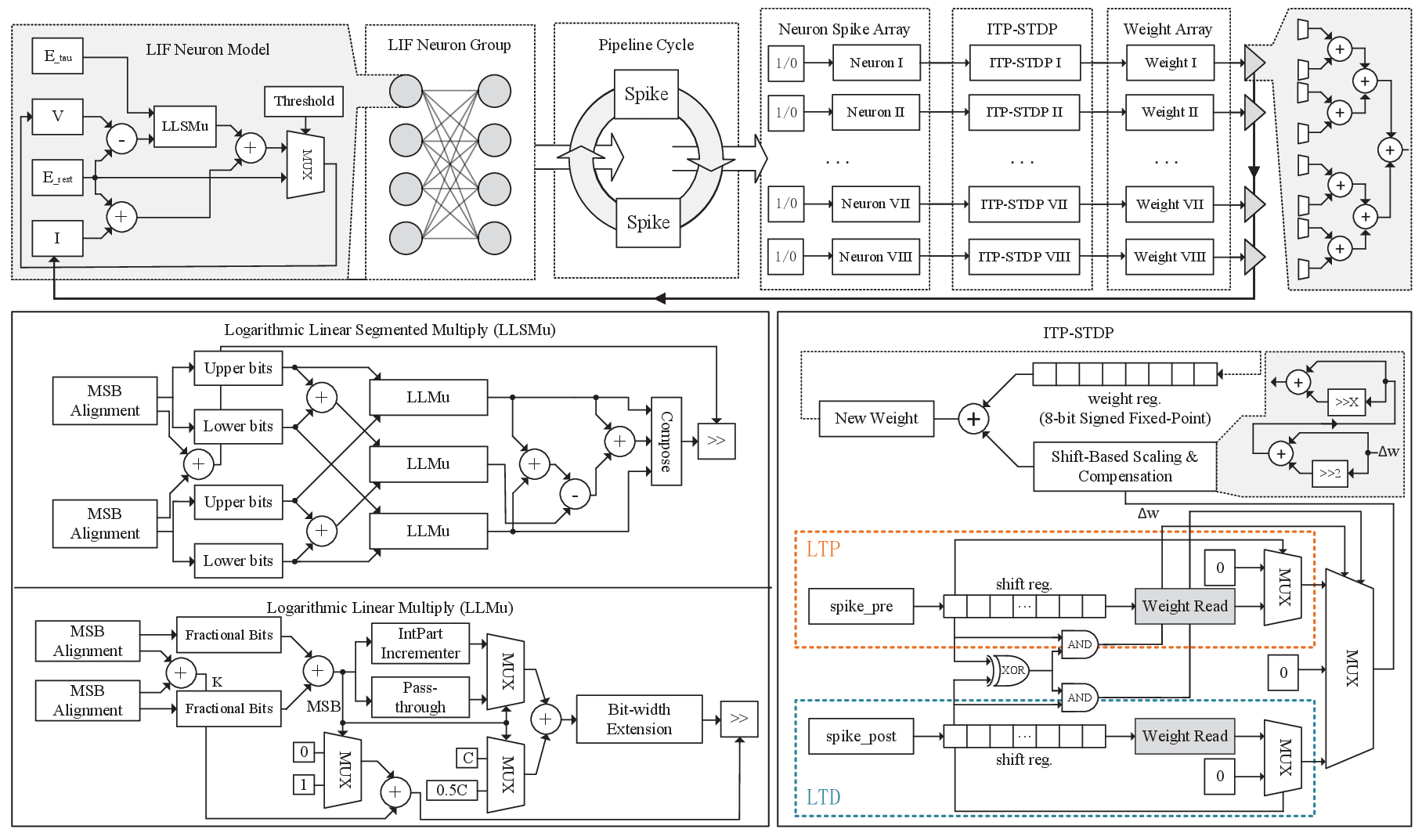}
    \caption{Block diagram of ITP-STDP learning engine hardware design.}
    \label{fig:hardware}
\end{figure*}

The ITP-STDP part in Fig. \ref{fig:hardware} illustrates the hardware architecture of the proposed ITP-STDP, which mainly consists of three parts. The first part is responsible for spike input and storage, forming a shift-register array that records spike history. The second part is responsible for weight reading and the corresponding weight-update computation in ITP-STDP. The third part is the control unit, which determines whether a weight update is triggered and whether the update is applied as LTP or LTD.

\begin{figure}[ht]
    \centering
    \begin{minipage}[t]{0.75\linewidth}
        \centering
        \includegraphics[width=1\textwidth]{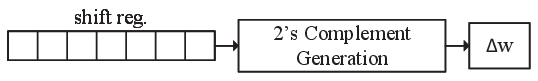}
        
        \scriptsize{(a) All-to-all spike-pairing.}
    \end{minipage}
    \begin{minipage}[t]{1\linewidth}
        \centering
        \includegraphics[width=1\textwidth]{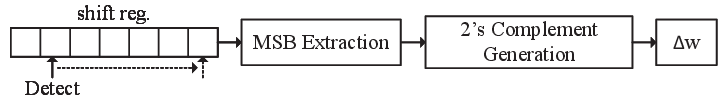}  
        
        \scriptsize{(b) Nearest-neighbor pairing.}
    \end{minipage}

    \caption{Block diagram of weight read module.}
    \label{fig:weright_read}
\end{figure}

\begin{figure}[ht]
    \centering
    \includegraphics[width=0.6\linewidth]{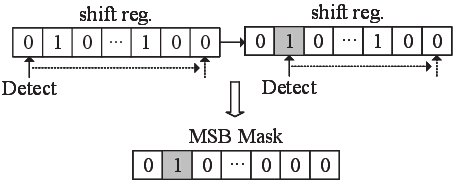}
    \caption{Block diagram of ITP-STDP hardware design.}
    \label{fig:mask}
\end{figure}

The spikes of two neurons are read in according to their pre- and post-synaptic relationship and stored in shift registers, which record the spike histories of the two neurons. These spike histories are then directly accessed for weight-update computation. Owing to the design of ITP-STDP, reading the spike history is equivalent to reading the weight variation itself. Different spike-pairing patterns lead to different weight-reading structures. Fig. \ref{fig:weright_read} (a) shows the weight-reading structure for all-to-all spike pairing, where the spike history is directly read and represented in two's-complement form as the weight update. By contrast, Fig. \ref{fig:weright_read} (b) illustrates the weight-reading structure for nearest-neighbor pairing. After the spike history is read, the one-hot representation of its most significant bit (MSB) is extracted. As illustrated in Fig. \ref{fig:mask}, the first `1` is detected from left to right, and only this bit is retained as `1` while all other bits are set to `0`, thereby forming the MSB mask. The resulting mask is then converted into two’s-complement form to obtain the intermediate weight-update value.

In the weight-update control logic, all control signals are generated from the spike activities of the two neurons. No weight update is triggered when both neurons either generate spikes or remain inactive. Once XOR and AND logic identify that only one neuron has fired, the resulting control signal drives a 3-input multiplexer to select between LTD and LTP. 
Subsequently, $\Delta w/2$ is obtained through a bit-shift operation and added to the original $\Delta w$, thereby realizing the $3/2$ compensation using only shift and addition operations. After compensation, the weight-scaling factor is applied using shift-and-add operations, yielding the final synaptic weight change. Finally, the resulting weight change is added to the current synaptic weight to obtain the updated weight, completing the weight-update process.

\subsection{Hardware Design of ITP-STDP Learning Engine}
Fig .\ref{fig:hardware} illustrates the proposed ITP-STDP learning engine structure. In this design, spikes are generated by th LIF neuron model and sequentially written into the corresponding neuron spike arrays to record spike histories. The LIF neuron model is constructed according to (\ref{eq.LIF}), where the time constant $\tau$ is directly provided as a constant parameter and multiplied by the difference between $v$ and $v_{rest}$. The multiplication unit adopts the approximate LLSMu multiplier to further reduce hardware resource usage and power consumption, and the corresponding LLSMu structure is also illustrated in Fig. \ref{fig:hardware}. The core operation of LLSMu is carried out by the LLM\_Mutiplier, while the compensation is computed in parallel with the multiplication process, thereby improving throughput, reducing computation latency, and decreasing computation error. The preprocessing module first detects the leading '1' and then shifts the entire operand so that the detected bit is aligned with the MSB. In addition, the LLSMu-based LIF neuron employs an eight-stage pipeline, which enables a single neuron hardware unit to serve as eight neuron models through time multiplexing. Their spikes are subsequently read out sequentially, one in each clock cycle.

After the spikes of all neurons are generated, they are written into the corresponding spike-history shift registers of ITP-STDP, and the STDP weight-update computation is then carried out. Once the synaptic weights of all connections are obtained, each post-synaptic neuron in the fully connected structure accumulates the weights from all pre-synaptic neurons as its input. The inclusion of each weight in the accumulation is directly controlled by the spike activity of the corresponding pre-synaptic neuron. The accumulated result is then taken as the input current $I$ and applied to the LIF neuron again. The proposed architecture is scalable, as larger network structures can be constructed by replicating and instantiating additional LIF neuron and STDP computation modules.

\section{Hardware Implementation Results}
\label{chap:6}
The ITP-STDP learning engine was implemented in Verilog HDL on a Xilinx Zynq UltraScale+ FPGA board, and all hardware designs were synthesized and evaluated using Vivado 2023.2. The corresponding performance metrics were obtained from the Vivado implementation results. In addition, these designs were implemented and evaluated at the typical process corner (1.00 V, $25\,^{\circ}\mathrm{C}$) in TSMC 28-nm technology. Synopsys Design Compiler, VCS, and PrimeTime PX were used to perform logic synthesis, dynamic timing simulation, and power analysis for all hardware implementations. For a fair comparison, the FPGA and ASIC results are reported separately. The standalone ITP-STDP module and the complete learning engine integrating the approximate LIF neuron are also implemented and evaluated separately. Furthermore, as SNNs are generally built by repeatedly replicating neuron models and learning algorithms, the evaluation and comparison of a single STDP algorithm with a simple learning-engine structure provide valuable insight into the scalability and overall performance of the network. In this context, optimizing the STDP learning algorithm and its hardware learning engine can effectively enhance the deployment performance of the complete network.

\subsection{FPGA Implementation Result of the proposed ITP-STDP}

\begin{table*}[ht]
    \centering
    \caption{Overall Comparison of FPGA Implementations of the ITP-STDP Learning Algorithm and Prior Designs}
    \resizebox{\textwidth}{!}{
    \begin{tabular}{cccccccc}
    \toprule \toprule
    \textbf{Model} & \textbf{C-STDP}\cite{jok17} & \textbf{C-STDP}\cite{jok19} & \textbf{P-STDP}\cite{lam19} & \textbf{P-STDP}\cite{lam19} & \textbf{P-STDP}\cite{nou18}& \textbf{t-STDP}\cite{x25} & \textbf{This Work}\\
    \midrule
    FPGA Platform & Virtex-6 & Virtex-6 & Spartan-6 & Spartan-6 & Spartan-6 & Zynq UltraScale+ & Zynq UltraScale+\\
    FPGA Technology & 40 nm & 40nm & 45 nm & 45 nm & 45 nm &  16 nm & 16 nm\\
    Bitwidth & 16-bit fixed-point & NR & 18-bit fixed-point & 18-bit fixed-point & NR &  8-bit fixed-point & 8-bit fixed-point\\
    Slice Registers & 292 & 139 & 642 & 12 & 45  & 114 & \textbf{8}\\
    Slice LUTs & 309 & 192 & 859 & 8 & 36  & 148 & \textbf{7}\\
    DSP Block & \textbf{0} & \textbf{0} & NR & NR & NR  & \textbf{0} & \textbf{0} \\
    Memory & NR & NR & NR & NR & NR &  \textbf{0} & \textbf{0}\\
    Max. Freq. & 322 MHz & 322 MHz & 362 MHz & 816 MHz & 186.32 MHz & 535.62 MHz & \textbf{2.73 GHz}\\
    Throughput & 644 MB/s & NR & 814.50 MB/s & 1.836 GB/s & NR &535.62 MB/s & \textbf{2.73 GB/s}\\
    Pipeline & 8 & NR & NR & NR & 3 & 5 & 2\\
    Latency & 24.84 ns & NR & NR & NR & 16.10 ns   & 9.335 ns &\textbf{0.732 ns}\\
    Power & NR & NR & 128 mW & 85 mW & NR & \textbf{$<$3 mW (100 MHz)} & \textbf{$<$ 3 mW (100 MHz)}\\
    Energy Consumption & NR & NR & 353.59 pJ/SOP & 104.17 pJ/SOP & NR  & 16.80 pJ/SOP & \textbf{4.03 pJ/SOP}\\
    Energy Efficiency & NR & NR & 2.83 GSOPS/W & 9.60 GSOPS/W & NR  & 59.51 GSOPS/W & \textbf{248.39 GSOPS/W}\\
    Error (Type) & NR & 0.336\% & 5.914\% & 6.13\% & 5.000\%  & 0.761\% & 0.49\% \\
     & (NR) & (NRMSD) & (NMSE) & (NMSE) & (RMSE)  & (NRMSD) & (NRMSD) \\
    \bottomrule
    \bottomrule
    \\[-2ex] 
    \end{tabular}}
    \label{tab:stdp_comp}
    \raggedright \scriptsize{NR means not reported; SOP means Spike Operations; C-STDP means calcium-based STDP; PSTDP means pair-based STDP; t-STDP means trace-STDP and t-STDP; The corresponding performance metrics were obtained from the Vivado implementation results.
    }
    \end{table*}

\subsubsection{Hardware Overhead Evaluation of the proposed ITP-STDP Learning Algorithm}
As shown in Table \ref{tab:stdp_comp}, with the same 8-bit weight width, ITP-STDP requires only 7.02\% of the slice registers and 2.70\% of the slice LUTs of LLMu t-STDP. Compared with \cite{lam19}, \cite{x25}, and \cite{jok17}--\cite{nou18}, ITP-STDP achieves the lowest slice register and slice LUT usage, corresponding to savings of 33.33\% to 98.75\% in slice registers and 12.50\% to 99.19\% in slice LUTs. This highlights the outstanding hardware resource efficiency of ITP-STDP. The substantial reduction in hardware utilization arises from the fact that ITP-STDP eliminates most of the weight-update computation itself, so that the majority of hardware overhead is only associated with spike and weight storage, together with shift operations. 

In addition, the higher hardware resource utilization in \cite{jok17} and \cite{jok19} can be partly attributed to the implementation of more complex calcium-based STDP variants. These designs also employ wider data bit widths than ITP-STDP, further increasing the hardware resource requirements. Moreover, the piecewise-linear (PWL) approximation introduces additional storage overhead for the segmented data. Implementing this storage with FPGA LUT resources instead of dedicated ROM further increases LUT utilization. Despite using an 18-bit data width, \cite{lam19} achieves low resource utilization by approximating model parameters with powers of two and limiting intermediate computations to 4-bit precision. These simplifications enable multiplication-free shift-based computation but result in relatively large approximation errors.
In summary, ITP-STDP achieves the lowest hardware resource utilization primarily because its power-of-two parameter approximation replaces multiplications with shift operations, while the intrinsic-timing mechanism avoids the arithmetic and storage overhead of PWL-based weight-update computation. The lightweight compensation further preserves low approximation error with minimal additional hardware cost.


\subsubsection{Running Speed Evaluation of the proposed ITP-STDP Learning Algorithm}
Among all the compared designs, the proposed ITP-STDP achieves the highest operating frequency, which is 3.35$\times$ to 14.65$\times$ higher than that of the other works. Moreover, although the weight width of ITP-STDP is only 8 bits, its throughput is still significantly improved. Compared with \cite{lam19}, \cite{x25}, and \cite{jok17}--\cite{nou18}, the throughput is increased by 1.49$\times$ to 5.10$\times$. Moreover, owing to its two-stage pipeline and high maximum operating frequency, ITP-STDP achieves the lowest latency among all compared designs, at only 0.732 ns, representing reductions of 92.16\%, 95.45\% and 97.05\% compared with \cite{x25}, \cite{nou18} and \cite{jok17}, respectively. This demonstrates that the proposed ITP-STDP achieves high operating speed while maintaining very low hardware resource consumption, thereby accelerating weight-update computation during SNN training. This improvement is mainly attributed to the intrinsic-timing mechanism of ITP-STDP, which directly derives the synaptic weight change from the spike history, eliminating explicit time-difference and weight-update computations. This substantially reduces the computational workload and thereby improves the operating speed.

\subsubsection{Energy Efficiency Evaluation of the proposed ITP-STDP Learning Algorithm}
The proposed ITP-STDP achieves significantly lower power consumption than \cite{lam19}, requiring only 2.34\%--3.53\% of the power consumed by those designs. At 100 MHz, its power consumption is identical to that of t-STDP, with both remaining below 3 mW. Nevertheless, when each design operates at its own maximum frequency, ITP-STDP reduces energy consumption by 76.01\% compared with t-STDP and by at least 96.13\% compared with the other works. As a result, ITP-STDP achieves the highest energy efficiency, exceeding the others by 4.17$\times$ to 87.77$\times$. The high energy efficiency of ITP-STDP mainly results from its low hardware resource utilization, which substantially reduces power consumption while maintaining a high operating frequency.

Overall, ITP-STDP combines low hardware resource utilization, high operating speed, and low energy consumption.
The low hardware resource utilization is attributed to the simplified timing representation and weight-update computation of ITP-STDP. Since spike events are stored sequentially in the spike-history register, the discrete timing difference is directly encoded by their relative bit positions, eliminating explicit timestamp storage and time-difference arithmetic. Moreover, the power-of-two learning kernel and the proposed compensation can be implemented using shifts and additions, avoiding transcendental-function units and multipliers while maintaining low approximation error.
Consequently, it is well suited for large-scale SNN implementation, where it can significantly reduce the required hardware resources, training time, and energy consumption.

\subsection{ASIC Implementation Result of the proposed ITP-STDP Learning Algorithm}

Table \ref{tab:STDP_ASIC} compares the ASIC implementation of the proposed ITP-STDP learning algorithm with representative state-of-the-art STDP designs, including CLSTDP, ImSTDP, and LLMu t-STDP. Among them, CLSTDP, ImSTDP, and LLMu t-STDP are all designed to implement the same STDP algorithm as ITP-STDP. With the same operand width and under similar process technologies, ITP-STDP achieves operating frequencies 8.70$\times$, 6.96$\times$, and 5.65$\times$ higher than those of CLSTDP, ImSTDP, and LLMu t-STDP, respectively. In addition, owing to its two-stage pipeline, the corresponding latency is reduced by 92.33\%, 90.42\%, and 92.92\%, while the throughput is increased by 21.47$\times$, 10.05$\times$, and 4.65$\times$, respectively. In addition, its area is only 0.84\%, 1.15\%, and 2.27\% of those of CLSTDP, ImSTDP, and LLMu t-STDP, respectively. Meanwhile, despite operating at a higher frequency, ITP-STDP still achieves the lowest power consumption, requiring only 5.91\%, 4.47\%, and 11.16\% of the corresponding values, respectively. As a result, ITP-STDP consumes the least energy per weight update and achieves the highest energy efficiency among all compared designs. Relative to prior designs, it reduces energy consumption by 98.10\% to 99.68\% and improves energy efficiency by 49.61$\times$ to 300.90$\times$.

\begin{table}
\centering
\caption{Overall Comparison of ASIC Implementations of the ITP-STDP Learning Algorithm and Prior Designs}
\resizebox{0.5\textwidth}{!}{
\begin{tabular}{c c c c c}
\toprule \toprule
\textbf{Method} & \textbf{CLSTDP}\cite{zha25} & \textbf{ImSTDP}\cite{zha25} & \textbf{LLMu t-STDP}\cite{x25} & \textbf{This Work}\\
\midrule
Node & 22 nm & 22 nm& 28 nm  & 28 nm\\
Voltage & NR & NR & 1.0 V  & 1.0 V\\
Bitwidth & 8-bit fixed-point & 8-bit fixed-point & 8-bit fixed-point  & 8-bit fixed-point\\
Frequency & 1.000 GHz & 1.250 GHz & 1.538 GHz & \textbf{8.696 GHz}\\
Latency & 3 ns & 2.40 ns &3.25& \textbf{0.23 ns} \\
Throughput & 0.387 GB/s & 0.787 GB/s & 1.538 GB/s & \textbf{8.696 GB/s} \\
Area & 1771.0 $\mu$m\textsuperscript{2} & 1294.5 $\mu$m\textsuperscript{2} & 655.074 $\mu$m\textsuperscript{2} & \textbf{29.61 $\mu$m\textsuperscript{2}}\\
Power & 1.68 mW & 1.33 mW & 0.89 mW & \textbf{0.09932 mW}\\
Energy Consumption & 3.453 pJ/SOP & 2.137 pJ/SOP & 0.579 pJ/SOP  & \textbf{0.011 pJ/SOP }\\
Energy Efficiency & 0.29 TSOPS/W & 0.47 TSOP/W & 1.73 TSOPS/W & \textbf{87.55 TSOPS/W}\\
Error (NRMSD) & \textbf{0} & 11.961\%* & 0.761\% & 0.49\% \\
Normalized FoM & $2.44\cdot10^{5}$ $\times$  & $1.00\cdot10^{5}$ $\times$ & $1.65\cdot10^{4}$ $\times$ & 1$\times$\\
\bottomrule
 \\[-2ex] 
\end{tabular}
}
\label{tab:STDP_ASIC}
\raggedright \scriptsize{NR means not reported; SOP means Spike Operations; * means not provided originally, but derived via reimplementation; FoM = Area $\times$ Latency $\times$ Energy Consumption $\div$ (1 $-$ Error)}
\end{table}

Moreover, after compensation, ITP-STDP incurs only 0.49\% computational error, whereas ImSTDP and LLMu t-STDP still exhibit higher errors, thereby affecting their biological plasticity. Therefore, to comprehensively evaluate the performance improvement of ITP-STDP, a composite figure of merit (FoM) is introduced. As shown by this metric, ITP-STDP improves the overall capability by $1.00\times10^{5}$ and $1.65\cdot10^{4}$ $\times$ compared with ImSTDP and LLMu t-STDP, respectively.Compared with CLSTDP using exact computation, it further achieves a $2.44 \times 10^{5}$-fold improvement. 

These improvements can be attributed to the simplified computation and data path of ITP-STDP. ImSTDP requires precomputed weight-update values to be stored and selected according to the calculated timing difference, introducing additional storage overhead, larger hardware area, and a longer computation path. These factors, together with its higher power consumption, result in lower energy efficiency. LLMu-based t-STDP primarily replaces multiplications with iterative shift-and-add operations without simplifying the underlying learning algorithm, leaving relatively high computational overhead and associated hardware and energy costs. Although it outperforms ImSTDP in terms of the composite FoM, its overall performance remains substantially lower than that of ITP-STDP.

\subsection{Hardware Implementation Results of ITP-STDP Learning Engine}

Furthermore, a hardware architecture based on the ITP-STDP learning engine, comprising eight LIF neurons arranged in a two-layer fully connected structure, was implemented and tested on both FPGA and ASIC platforms. The corresponding results are presented in Table \ref{tab:engine}.

\begin{table*}[!htbp]
    \centering
    \caption{Hardware implementation results of the proposed ITP-STDP learning engine}
    \label{tab:hardware_results}
    \renewcommand{\arraystretch}{1.2}
    \setlength{\tabcolsep}{4pt}
    \resizebox{\textwidth}{!}{
    \begin{tabular}{c c c c c c  c c c c c c}
    \toprule \toprule
        \multirow{2}{*}{\textbf{Platform}} 
        & \multirow{2}{*}{\textbf{Neuron Num.}}
        & \multirow{2}{*}{\textbf{Synapse Num.}}
        & \multicolumn{2}{c}{\textbf{Bitwidth}}
        & \multicolumn{2}{c}{\textbf{Area}}
        & \multirow{2}{*}{\textbf{Max. Frequency}}
        & \multirow{2}{*}{\textbf{Latency}}
        & \multirow{2}{*}{\textbf{Power}}
        & \multirow{2}{*}{\textbf{Power Consumption}}
        & \multirow{2}{*}{\textbf{Energy Efficiency}} \\
        \cmidrule(lr){6-7} 
        \cmidrule(lr){4-5}
        & &  & Neuron & Weight & LUT & Reg. & & & & & \\
    \midrule
        \textbf{Zynq UltraScale+} 
        & 8 & 16  & 8 & 8 & 387 & 251 & 505.82 MHz & 19.77 ns & 17 mW & 33.61 pJ/SOP & 29.75 GSOPS/W \\

        \textbf{TSMC’s 28 nm tech.} 
        & 8 & 16 & 8 & 8 & \multicolumn{2}{c}{2127.76 $\mu$m\textsuperscript{2}} & 2.41 GHz & 4.15 ns & 3.68 mW & 1.53 pJ/SOP & 654.79 GSOPS/W \\
    \bottomrule
    \bottomrule
        \\[-2ex] 
    \end{tabular}
    }
    \label{tab:engine}
    \raggedright \scriptsize{SOP means Spike Operations}
\end{table*}

By combining an LLSMu-based LIF neuron model with ITP-STDP and directly accessing weights from registers, the proposed design avoids complex inter-neuron connection circuitry, thereby significantly reducing the hardware resource utilization of the ITP-STDP learning engine. On the FPGA platform, it occupies only 387 slice LUTs and 251 slice registers, without requiring any additional hardware resources such as DSPs or memory blocks. On the ASIC platform implemented in 28-nm technology, its chip area is only $2127.76~\mu\mathrm{m}^2$. Moreover, because the computationally intensive operations in both the neuron model and the STDP algorithm are optimized and implemented with a pipelined structure, the design achieves a very high operating frequency, reaching 505.82 MHz on FPGA and 2.41 GHz on ASIC. However, ITP-STDP employs a 10-stage pipeline structure, including 8 stages for the LIF neuron model and 2 stages for ITP-STDP, which results in latencies of 19.77 ns and 4.15 ns on FPGA and ASIC, respectively. Nevertheless, owing to its very high operating frequency, the overall latency remains at a relatively low level.

Notably, the ITP-STDP learning engine also demonstrates excellent power efficiency. At its maximum operating frequency, the power consumption is only 11 mW on FPGA and 3.68 mW on ASIC. Owing to its high operating frequency, each spike operation requires only 33.61 pJ and 1.53 pJ, respectively, resulting in energy efficiencies as high as 29.75 GSOP/W and 654.79 GSOP/W. This implies that large-scale SNNs built with the ITP-STDP learning engine can be deployed with very low hardware resource usage while maintaining high-speed computation and extremely low energy consumption.

\section{Conclusion}
\label{chap:7}
In this paper, a novel Intrinsic-Timing Power-of-Two STDP (ITP-STDP) algorithm is presented for spiking neural networks (SNNs), together with the hardware implementation of an on-chip learning engine integrating the proposed ITP-STDP algorithm and an approximate LIF neuron model.
A mean-field synaptic drift model was further developed to analyze the dynamics of ITP-STDP. The proposed ITP-STDP was also evaluated using three SNN architectures of different scales across multiple datasets. The results demonstrate that, after error compensation, ITP-STDP preserves dynamical behavior that is highly consistent with conventional STDP, while remaining compatible with different network architectures without compromising training or inference accuracy. The proposed ITP-STDP learning algorithm and ITP-STDP learning engine were further validated through FPGA and ASIC implementations, demonstrating low hardware resource utilization, high operating frequency and throughput, and substantial power-efficiency advantages. These advantages make the ITP-STDP learning engine particularly suitable for large-scale SNN deployment, as it substantially reduces energy consumption and hardware resource requirements while enabling efficient implementation on edge devices.



 




\vfill


\begin{thebibliography}{00}

\bibitem{joo22} B. Joo , J.-W. Han, and B.-S. Kong, “Energy- and Area-Efficient CMOS Synapse and Neuron for Spiking Neural Networks With STDP Learning,” \textit{IEEE Trans. Circuits Syst. I-Regul. Pap.}, vol. 69, no. 9, pp. 3632-3642, Sept. 2022.

\bibitem{liu22} Y. Liu \textit{et al.}, “FPGA-NHAP: A General FPGA-Based Neuromorphic Hardware Acceleration Platform With High Speed and Low Power,” \textit{IEEE Trans. Circuits Syst. I-Regul. Pap.}, vol. 69, no. 6, pp. 2553-2566, Mar. 2022.

\bibitem{maa97} W. Maass, “Networks of spiking neurons: The third generation of neural network models,” \textit{Neural Netw.}, vol. 10, no. 9, pp. 1659-1671, Dec. 1997.

\bibitem{che23} J. Chen , N. Skatchkovsky, and O. Simeone, “Neuromorphic Integrated Sensing and Communications,” \textit{IEEE Wirel. Commun. Lett.}, vol. 12, no. 3, pp. 476-480, Mar. 2023.

\bibitem{nic13} E. Nichols, L. J. McDaid, and N. Siddique, “Biologically Inspired SNN for Robot Control,” \textit{IEEE T. Cybern.}, vol. 43, no. 1, pp. 115-128, Feb. 2013.

\bibitem{eqp25} G. Eappen \textit{et al.}, “Neuromorphic Models for Energy-Efficient Onboard Interference Detection in Satellite Systems,” \textit{IEEE Trans. Aerosp. Electron. Syst.}, vol. 61, no. 6, pp. 17682-17702, Dec. 2025.

\bibitem{liu24} W. Liu \textit{et al.}, “SC-PLR: An Approximate Spiking Neural Network Accelerator with On-Chip Predictive Learning Rule,” \textit{IEEE Trans. Biomed. Circuits Syst.}, vol. 18, no. 5, pp. 1156-1165, Oct. 2024.

\bibitem{wan22} H. Wang \textit{et al.}, “TripleBrain: A Compact Neuromorphic Hardware Core with Fast On-Chip Self-Organizing and Reinforcement Spike-Timing Dependent Plasticity,” \textit{IEEE Trans. Biomed. Circuits Syst.}, vol. 16, no. 4, pp. 636-650, Aug. 2022.

\bibitem{fre19} C. Frenkel \textit{et al.}, “A 0.086-mm2 12.7-pJ/SOP 64k-Synapse 256-Neuron Online-Learning Digital Spiking Neuromorphic Processor in 28-nm CMOS,” \textit{IEEE Trans. Biomed. Circuits Syst.}, vol. 13, no. 1, pp. 145–158, Feb. 2019.

\bibitem{kim24} J. Kim \textit{et al.}, “Hardware-Efficient Emulation of Leaky Integrate-and-Fire Model Using Template-Scaling-Based Exponential Function Approximation,” \textit{IEEE Trans. Circuits Syst. I-Regul. Pap.}, vol. 68, no. 1, pp. 350-362, Oct. 2024.

\bibitem{dav18} M. Davies \textit{et al.}, “Loihi: A neuromorphic manycore processor with onchip learning,” \textit{IEEE Micro}, vol. 38, no. 1, pp. 82-99, Jan. 2018.

\bibitem{li23} J. Li \textit{et al.}, “FireFly: A High-Throughput Hardware Accelerator for Spiking Neural Networks With Efficient DSP and Memory Optimization,” \textit{IEEE Trans. Very Large Scale Integr. (VLSI) Syst.}, vol. 31, no. 8, pp. 1178-1191, Aug. 2023.

\bibitem{sad25} M. Sadeghi \textit{et al.}, “NEXUS: A 28nm 3.3pJ/SOP 16-Core Spiking Neural Network with a Diamond Topology for Real-Time Data Processing,” \textit{IEEE Trans. Biomed. Circuits Syst.}, vol. 19, no. 3, pp. 523-535, June 2025.

\bibitem{xia25} H. Xia \textit{et al.}, “A High Accuracy and Hardware Efficient Approximate Computing based Leaky integrate-and-fire Neuron Model,” in \textit{Proc. IEEE 38th Int. Syst.-on-Chip Conf. (SOCC)}, Oct. 2025, pp. 1-6.

\bibitem{lob19} J. L. Lobo \textit{et al.}, “Spiking Neural Networks and Online Learning: An Overview and Perspectives,” \textit{Neural Netw.}, vol. 121, pp. 88-100, July 2019.

\bibitem{heb05} D. O. Hebb, \textit{The Organization of Behavior: A Neuropsychological Theory.} New York, NY, USA: Wiley, 2005.



\bibitem{jia25} Y. Jiang \textit{et al.}, “Circuit Modeling and Analysis of Memristor Crossbar Array-Based SNN With Spike-Mapping STDP Implementation,” \textit{IEEE Trans. Very Large Scale Integr. (VLSI) Syst.}, vol. 34, no. 1, pp. 193-204, Oct. 2025.



\bibitem{son00} S. Song, K. D. Miller, and L. F. Abbott, “Competitive Hebbian learning through spike-timing-dependent synaptic plasticity,” \textit{Nature Neurosci.}, vol. 3, no. 9, pp. 919–926, Sept. 2000.



\bibitem{voh23} S. K. Vohra \textit{et al.}, “Full CMOS Circuit for Brain-Inspired Associative Memory With On-Chip Trainable Memristive STDP Synapse,” \textit{IEEE Trans. Very Large Scale Integr. (VLSI) Syst.}, vol. 31, no. 7, pp. 993-1003, Jul. 2023.


\bibitem{fre22} C. Frenkel and G. Indiveri, “ReckOn: A 28nm Sub-mm2 Task-Agnostic Spiking Recurrent Neural Network Processor Enabling On-Chip Learning over Second-Long Timescales,” in \textit{IEEE Int. Solid-State Circuits Conf. (ISSCC) Dig. Tech. Papers}, vol. 65, Feb. 2022, pp. 1-3.



\bibitem{liu24} Y. Liu \textit{et al.}, “Implementation of Multiple-Step Quantized STDP Based on Novel Memristive Synapses,” \textit{IEEE Trans. Very Large Scale Integr. (VLSI) Syst.}, vol. 32, no. 8, pp. 1369-1379, Aug. 2024.


\bibitem{li21} S. Li \textit{et al.}, “A Fast and Energy-Efficient SNN Processor with Adaptive Clock/Event-Driven Computation Scheme and Online Learning,” \textit{IEEE Trans. Circuits Syst. I-Regul. Pap.}, vol. 68, no. 4, pp. 1543-1552, Apr. 2021.

\bibitem{nei14} D. Neil and S.-C. Liu, “Minitaur, an event-driven FPGA-based spiking network accelerator,” \textit{IEEE Trans. Very Large Scale Integr. (VLSI) Syst.}, vol. 22, no. 12, pp. 2621–2628, Dec. 2014.

\bibitem{lam19} C. Lammie \textit{et al.}, “Efficient FPGA Implementations of Pair and Triplet-Based STDP for Neuromorphic Architectures,” \textit{IEEE Trans. Circuits Syst. I-Regul. Pap.}, vol. 66, no. 4, pp. 1206-1219, Apr. 2019.

\bibitem{nou18} M. Nouri \textit{et al.}, “A digital neuromorphic realization of pair-based and triplet-based spike-timing-dependent synaptic plasticity,” \textit{IEEE Trans. Circuits Syst., II, Exp. Briefs}, vol. 65, no. 6, pp. 804–808, Jun. 2018.

\bibitem{zha25} D. Zhao \textit{et al.}, “ImSTDP: Implicit Timing On-Chip STDP Learning,” \textit{IEEE Trans. Circuits Syst. I-Regul. Pap.}, vol. 72, no. 2, pp. 868-881, Feb. 2025.

\bibitem{sun22} C. Sun \textit{et al.}, “An Energy Efficient STDP-Based SNN Architecture With On-Chip Learning,” \textit{IEEE Trans. Circuits Syst. I-Regul. Pap.}, vol. 69, no. 12, pp. 5147-5158, Dec. 2022.



\bibitem{bai20} T. J. Bailey \textit{et al.}, “Development of a Short-Term to Long-Term Supervised Spiking Neural Network Processor,” \textit{IEEE Trans. Very Large Scale Integr. (VLSI) Syst.}, vol. 28, no. 11, pp. 2410–2423, Nov. 2020.


\bibitem{par20} J. Park and S.-D. Jung, “Presynaptic Spike-Driven Spike Timing-Dependent Plasticity with Address Event Representation for Large-Scale Neuromorphic Systems,” \textit{IEEE Trans. Circuits Syst. I-Regul. Pap.}, vol. 67, no. 6, pp. 1936-1947, Jun. 2020.

\bibitem{gom18} S. Gomar and M. Ahmadi, "Digital Realization of PSTDP and TSTDP Learning," in \textit{Proc. Int. Joint Conf. Neural Netw. (IJCNN)}, Jul. 2018, pp. 1-5.

\bibitem{zho23} Y. Zhong \textit{et al.}, “An Efficient Neuromorphic Implementation of Temporal Coding-Based On-Chip STDP Learning,” \textit{IEEE Trans. Circuits Syst., II, Exp. Briefs}, vol. 70, no. 11, pp. 4241-4245, Nov. 2023.

\bibitem{wu21} J. Wu \textit{et al.}, “Efﬁcient Design of Spiking Neural Network With STDP Learning Based on Fast CORDIC,” \textit{IEEE Trans. Circuits Syst. I-Regul. Pap.}, vol. 68, no. 6, pp. 2522-2534, June 2021.

\bibitem{x25} H. Xia \textit{et al.}, “An Approximate Computing-Based Spiking Neural Networks Neuron Model and STDP Learning,” \textit{IEEE Trans. Circuits Syst. I-Regul. Pap.}, vol. 73, no. 4, pp. 2621–2634, Oct. 2025.

\bibitem{mit62} J. N. Mitchell, “Computer multiplication and division using binary  logarithms,” \textit{IRE Trans. Electron. Comput.}, vol. 11, no. 4, pp. 512–517, Aug. 1962.

\bibitem{kar63} A. A. Karatsuba and Y. Ofman, “Multiplication of multidigit numbers  on automata,” \textit{Sov. Phys. Doklady}, vol. 7, pp. 595–596, Dec. 1963.

\bibitem{saa18} H. Saadat, H. Bokhari, and S. Parameswaran, “Minimally biased multipliers for approximate integer and floating-point multiplication,” \textit{IEEE Trans. Comput.-Aided Design Integr. Circuits Syst.}, vol. 37, no. 11, pp. 2623–2635, Nov. 2018.

\bibitem{bur04} A. N. Burkitt \textit{et al.}, “Spike-Timing-Dependent Plasticity: The Relationship to Rate-Based Learning for Models with Weight Dynamics Determined by a Stable Fixed Point,” \textit{Neural Comput}, vol. 16, no. 5, pp. 885-940, May 2004.

\bibitem{xi25} H. Xia \textit{et al.}, "Convolutional \& Spiking Neural Networks for the Efficient Classification of Rotor Faults in Induction Motors via Stator Current and Stray Flux Signals," in \textit{Proc. 15th IEEE Int. Symp. Diag. Elect. Mach., Power Electron. Drives}, Aug. 2025, pp. 1-7.


\bibitem{xia26} H. Xia \textit{et al.}, “Efficient \& Lightweight Classification of Rotor Bar Faults in Induction Motors by Convolutional and Spiking Neural Networks,” \textit{IEEE Trans. Ind. Appl.}, Early Access, pp. 1-13, Jun. 2026.

\bibitem{wan23} K. Wang \textit{et al.}, “Comparison and Selection of Spike Encoding Algorithms for SNN on FPGA,” \textit{IEEE Trans. Biomed. Circuits Syst.}, vol. 17, no. 2, pp. 129-141, Feb. 2023.

\bibitem{ami17} A. Amir \textit{et al.}, "A Low Power, Fully Event-Based Gesture Recognition System," in \textit{ Proc. IEEE Conf. Comput. Vis. Pattern Recognit.}, Jul. 2017, pp. 7243-7252.

\bibitem{jok17} E. Jokar and H. Soleimani, “Digital Multiplierless Realization of a Calcium-Based Plasticity Model,” \textit{IEEE Trans. Circuits Syst. II-Express Briefs}, vol. 64, no. 7, pp. 832-836, July 2017

\bibitem{jok19} E. Jokar, H. Abolfathi and A. Ahmadi , “A Novel Nonlinear Function Evaluation Approach for Efficient FPGA Mapping of Neuron and Synaptic Plasticity Models,” \textit{IEEE Trans. Biomed. Circuits Syst.}, vol. 13, no. 2, pp. 454-469, Apr. 2019.

\bibitem{nou18} M. Nouri \textit{et al.}, “A Digital Neuromorphic Realization of Pair-Based and Triplet-Based Spike-Timing-Dependent Synaptic Plasticity,” \textit{IEEE Trans. Circuits Syst. II-Express Briefs}, vol. 65, no. 6, pp. 804-808, Jun. 2018.




\end{thebibliography}
\end{document}